\documentclass[a4paper,fleqn]{cas-sc}

\usepackage[numbers, sort&compress]{natbib}

\usepackage[T1]{fontenc}
\usepackage{cmap}
\usepackage{amssymb}
\usepackage{bm}
\usepackage{graphicx}
\usepackage[utf8]{inputenc}
\usepackage{color}
\usepackage{subfigure}
\usepackage{amsmath, bm}
\usepackage{makecell}
\usepackage{graphicx,xcolor,framed}
\usepackage{amsmath}
\usepackage{braket}
\usepackage{ragged2e}
\usepackage{verbatim}
\usepackage[numbers,sort&compress]{natbib}
\usepackage{CJKutf8}
\usepackage{newunicodechar}
\newunicodechar{́}{\'{}}

\def\tsc#1{\csdef{#1}{\textsc{\lowercase{#1}}\xspace}}
\tsc{WGM}
\tsc{QE}
\begin{document}
\begin{CJK}{UTF8}{gbsn}
\let\WriteBookmarks\relax
\def\floatpagepagefraction{1}
\def\textpagefraction{.001}

\shorttitle{Progress in Quantum Electronics}    

\shortauthors{Yulin Shen et al.}  

\title [mode = title]{Nonlinear photocurrent in quantum materials for broadband photodetection}  

%\tnotemark[<tnote number>] 

%\tnotetext[<tnote number>]{<tnote text>} 


\author[1]{Yulin Shen}

\author[1]{Louis Primeau}

\author[1]{Jiangxu Li}

\author[4]{Tuan-Dung Nguyen}

\author[1,2,3]{David Mandrus}

\author[4]{Yuxuan Cosmi Lin}

\author[1,5]{Yang Zhang}



%\ead{<email address>}

%\ead[url]{<URL>}

%\credit{<Credit authorship details>}

\affiliation[1]{organization={Department of Physics and Astronomy, University of Tennessee, Knoxville},
%          citysep={}, % Uncomment if no comma needed between city and postcode
            postcode={TN 37996}, 
            country={USA}}

\affiliation[2]{organization={Department of Materials Sciences and Engineering, University of Tennessee, Knoxville},
%          citysep={}, % Uncomment if no comma needed between city and postcode
            postcode={TN 37996}, 
            country={USA}}

\affiliation[3]{organization={Materials Science and Technology Division, Oak Ridge National Laboratory},
%          citysep={}, % Uncomment if no comma needed between city and postcode
            postcode={TN 37831}, 
            country={USA}}

\affiliation[4]{organization={Department of Materials Science and Engineering, Texas A\&M University, College Station},
%          citysep={}, % Uncomment if no comma needed between city and postcode
            state={TX},
            country={USA}}

\affiliation[5]{organization={Min H. Kao Department of Electrical Engineering and Computer Science, University of Tennessee, Knoxville},
%          citysep={}, % Uncomment if no comma needed between city and postcode
            postcode={TN 37996}, 
            country={USA}}




\begin{abstract}
Unlocking the vast potential of optical sensing technology has long been hindered by the challenges of achieving fast, sensitive, and broadband photodetection at ambient temperatures. In this review, we summarize recent progress in the study of nonlinear photocurrent in topological quantum materials, and its application in broadband photodetection without the use of p-n junction based semiconductor diodes. The intrinsic quadratic transverse current-input voltage relation is used to rectify the alternating electric field from incident radio, terahertz or infrared waves into a direct current, without a bias voltage and at zero magnetic field. We review novel photocurrents in several material systems, including topological Weyl semimetals, chiral crystals, ferroelectric materials, and low dimensional topological insulators. These quantum materials hold tremendous promise for broadband high-frequency rectification and photo-detection, featuring substantial responsivity and detectivity.

\end{abstract}


\maketitle

% Main text
\section{Introduction}\label{Introduction}

Nonlinear photocurrent refers to a quadratic dc current in response to the ac external electric field from light radiation~\cite{moore2010confinement}. It serves as a key observable in many condensed matter phenomena, such as sensing of electric charge modulation~\cite{yang2022spectroscopy, song2015energy}, spin order ~\cite{ganichev2003spin, ganichev2014interplay}, collective excitations~\cite{lundeberg2017thermoelectric, barati2017hot}, quantum geometry, and quantum kinetic processes in quantum materials. The development of high-performance photocurrent devices using quantum materials advances our fundamental understanding of the various photocurrent mechanisms by harnessing photocurrent as a widely accepted diagnostic tool~\cite{ma2023photocurrent}. In this review, we focus on the theory and application of nonlinear photocurrent in broadband photodetection, spanning from visible through infrared (IR) to terahertz (THz) wavelengths. 

The recently discovered nonlinear photocurrent in topological quantum materials is characterized by a response voltage (or current) quadratically dependent on the external electric field. The resulting quadratic V-I relation~(Fig\ref{fig2}. d) leads to possible applications in high frequency rectification and photodetection. There are three types of nonlinear photocurrent widely discussed in quantum materials: the nonlinear Hall effect below infrared frequencies, and the linear and circular photogalvanic effect above infrared frequencies. All three of these processes require the materials to be noncentrosymmetric, and are closely related to the geometric properties of the quantum wavefunction~\cite{wang2020electrically, heidari2022nonlinear, tanaka2023nonlinear}. In this review, we focus on topological Weyl semimetals~\cite{armitage2018weyl, weng2015weyl, huang2015weyl, lv2015experimental, xu2015discovery, yang2015weyl, bradlyn2016beyond, tang2017multiple, chang2017unconventional, rao2019observation, sanchez2019topological, takane2019observation}, topological chiral crystals~\cite{chang2018topological}, ferroelectric materials~\cite{rehman2020topology}, and magnetic materials\cite{gao2023quantum}, which lack inversion symmetry and host topologically non-trivial band structures in the bulk or on the surface, with a focus on 2D materials. 

Topological quantum materials stand out as excellent material candidates for photodetection due to several compelling reasons:~(1) Topological materials are abundant in earth, providing rich raw materials for device fabrication.~(2) First principle simulations and machine-learning are powerful tools for advancing scientific discovery in new topological materials~\cite{wang2021machine, ma2023topogivity, tang2019comprehensive, vergniory2019complete, zhang2019catalogue, yeung2020elucidating, fuchs2020se, gao2020deep, lu2020extracting} and it is possible to find more effective materials for photodetection via computational modeling.~(3) Molecular-beam epitaxy~(MBE) and metalorganic chemical vapor deposition~(MOCVD) are effective in producing large-area topological films~\cite{tokumitsu1989photo, bhuiyan2023tutorial, tang2022high}, benefiting device fabrication.

This review intends to provide a comprehensive overview of advancements in broadband photodetection through nonlinear photocurrent. We start by presenting the basic theory of conventional methods and their limitations, followed by the theory of and recent experiments in the nonlinear photocurrent in quantum materials , and its application and related advantages in broadband photodetection over conventional methods. Looking forward, we outline crucial challenges for further development, encompassing fundamental physics concepts and device fabrication. These areas not only serve as focal points for current research but also represent fertile ground for further exploration of accessible physics and hold promise for future technological applications.

\section{Conventional photodetection methods}

Conventional photodetection approaches typically hinge on photodiodes(Fig\ref{fig2}. b), which are based on p-n junction~(Fig\ref{fig2}. a)~\cite{kozawa2016photodetection}, and thermal bolometers~\cite{varpula2021nano}. A p-n junction consists of two regions of semiconducting material, one doped with an excess of electrons~(n-type) and the other with a deficiency of electrons~(p-type). The interface of these two regions is the junction, where a depletion layer forms due to diffusion of charge carriers. The resulting charge accumulation creates an electric field that opposes further diffusion, leading to a built-in potential difference across the junction. The current is generated by the diffusion of the carrier. The photodiode is a semiconductor device with a p-n junction structure. In a photodiode, when an incident photon's energy exceeds the semiconductor bandgap, the electron in the valence band is excited to the conduction band, generating an electron-hole pair. Due to diffusion of carriers at the junction, a depletion region is formed, accompanied by an electric field which causes the electron-hole pair to separate. The oppositely moving electron and hole lead to photocurrent. When reverse biased, there will be an enlarged depletion region, increasing the responsive area and increasing the acceleration of the electrons. The bias also leads to increased dark current when there is no incident light. Note that the applied bias can only be reverse bias in a photodiode as under forward bias, the response current is nonlinear with the light intensity so that the detection is much less sensitive. Photodiodes have a few limitations in photodetection~\cite{weckler1967operation}:~(1) limited spectral range that is determined by the materials' bandgap. For example, silicon-based photodiode only detect light up to 1.1 $\mu$m wavelength, while germanium-based photodiode can only detect light up to 1.8 $\mu$m~\cite{kozawa2016photodetection, wei2017photodetectors}.~(2) When the incident light intensity is too high, the photocurrent will saturate, and no longer linearly proportional to the light intensity. This is because when the light intensity is low, the number of photons received is proportional to the light intensity and each photon generates an electron-hole pair, leading to linear response. When the light intensity is very high, the number of electron-hole pairs exceeds the capacity of the photodiode to separate and collect as the built-in electric field in the depletion region has a finite strength. Once the saturation point is reached, the photodiode does not conduct additional current. Although this can be avoided by controlling the light intensity and reverse bias voltage, the performance and the power consumption of device will be affected~\cite{malik20222d}. 

\begin{figure*}[pos=!h]
	\centering
		\includegraphics[width=1\textwidth]{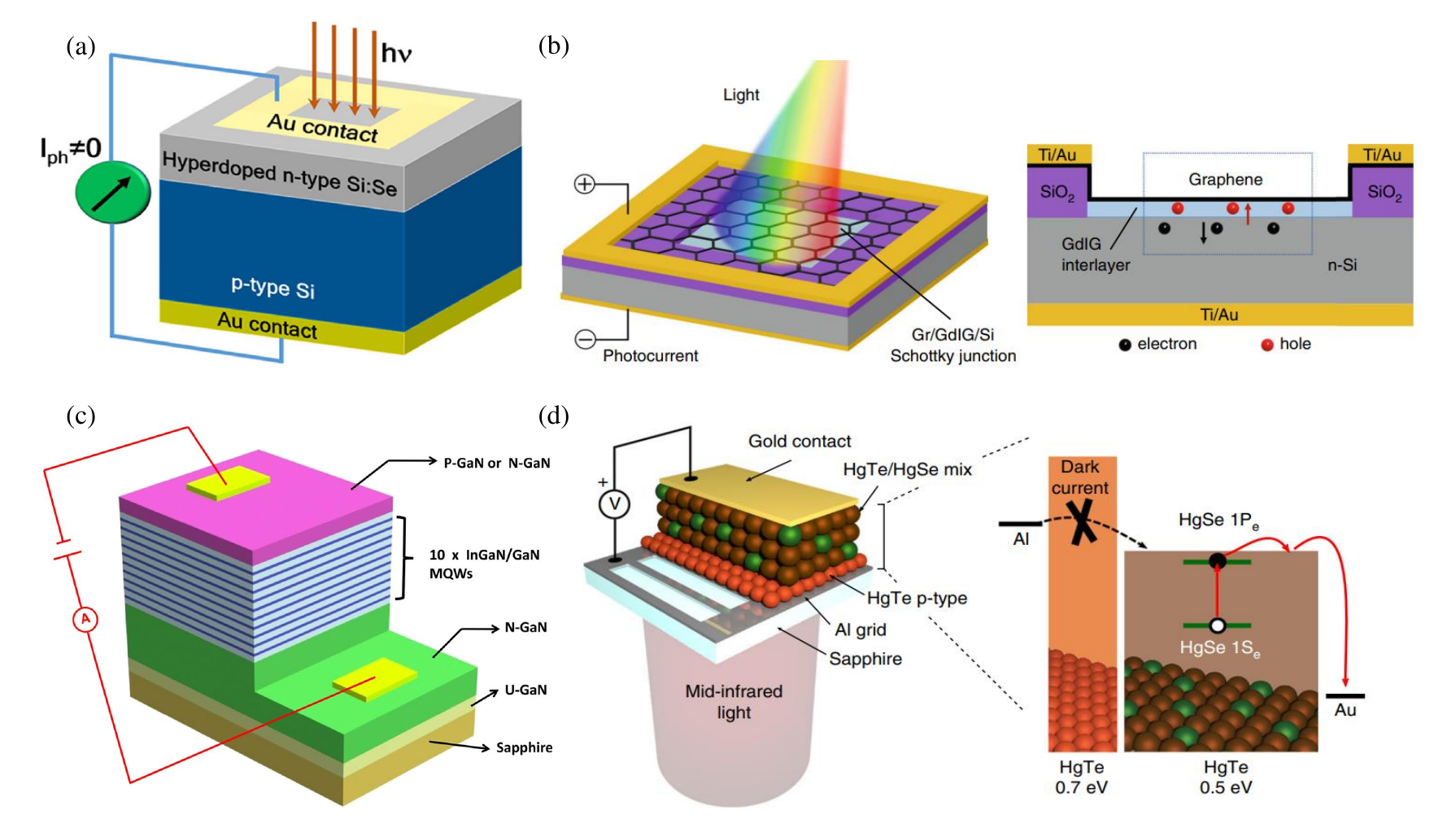}
	  \caption{\textbf{Different types of photodetectors~\cite{rogalski2002infrared}.} \textbf{a}, Schematic diagram of an impurity absorption photodetector based on Si hyperdoped with Se concentrations \cite{berencen2017room}. Its advantages include that it has very long wavelength operation and it depends on relatively simple technology. Its disadvantages are it has high thermal generation and it operates at extremely low temperature.\textbf{b}, Schematic diagram of a graphene/silicon Schottky barriers photodetector \cite{ji2022high}. Its advantages include that it is low-cost and high yields. Its disadvantages are it has low quantum efficiency and operates at low temperature. \textbf{c}, Schematic diagram of an InGaN/GaN quantum wells photodetector \cite{yang2017enhanced}. Its advantages include that it is a multicolor detector and it has good uniformity over large area. \textbf{d}, Schematic diagram of a colloidal quantum dot photodetector \cite{livache2019colloidal}. Its advantage includes that it has low thermal generation. Its disadvantage is it is complicated to design and growth. Panel \textbf{a} adapted with permission from \cite{berencen2017room}, panel \textbf{b} adapted with permission from \cite{ji2022high}, panel \textbf{c} adapted with permission from \cite{yang2017enhanced} and panel \textbf{d} adapted with permission from \cite{livache2019colloidal}}\label{fig1}
\end{figure*}

\begin{figure*}[pos=!h]
	\centering
		\includegraphics[width=1\textwidth]{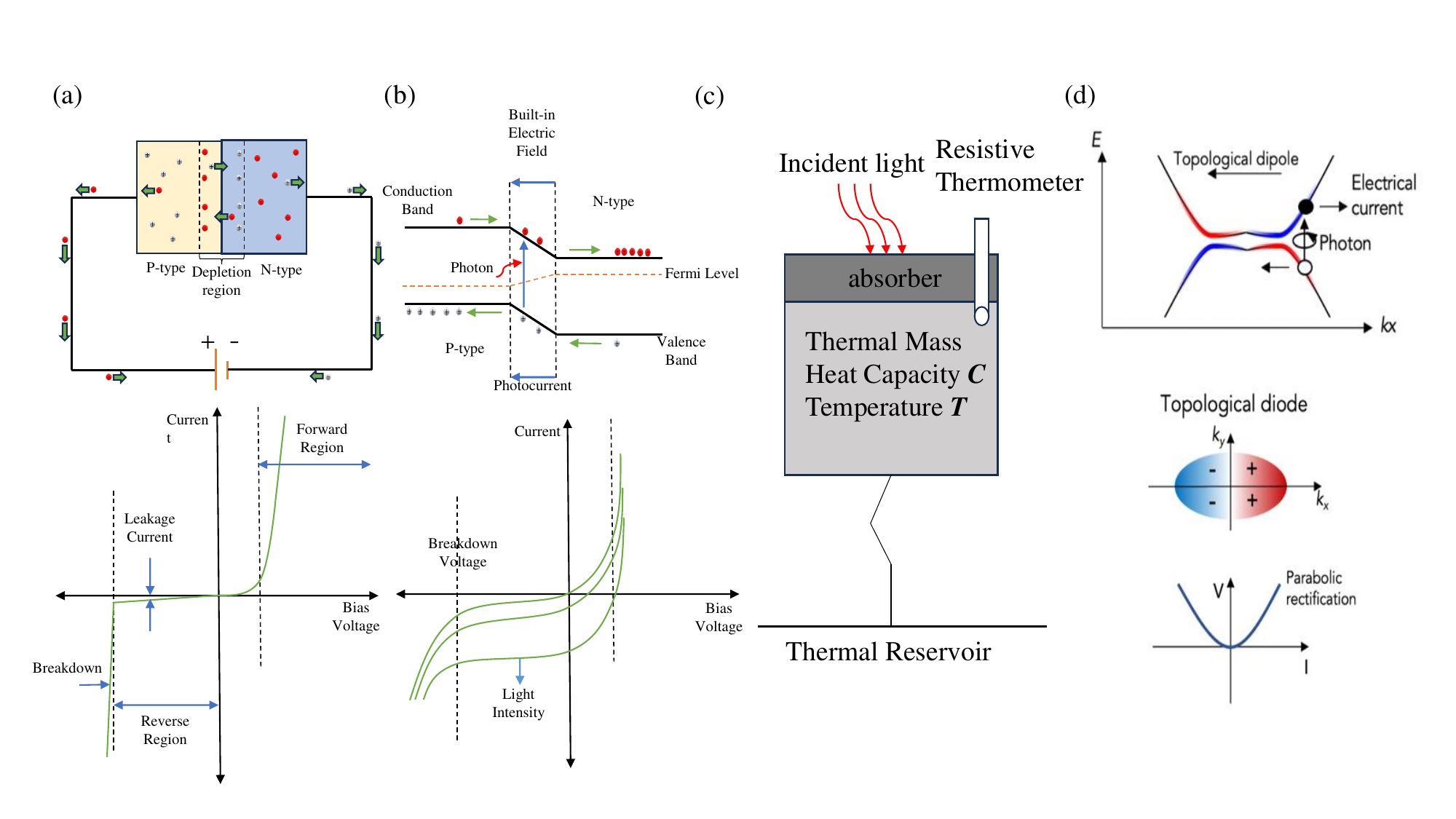}
	  \caption{\textbf{Conventional and new mechanics for photodetection.} \textbf{a}, Schematic diagram of p-n junction. \textbf{b}, Photodiodes that convert light into an electrical current. The DC current is generated when photons are absorbed in the photodiode, and the minimum operating photon energy is set by the semiconductor bandgap. \textbf{c}, Schematic diagram of thermal bolometer. Incident radiation heats the well-characterized absorber, causing a change in resistance, which is what is actually measured. \textbf{d}, The mechanism of the unique topological rectification.}\label{fig2}
\end{figure*}

The thermal bolometer operates on the photothermal effect, which converts photon energy into a heat gradient, and then measured by a thermometer. Its structure and operating principle are shown in Fig\ref{fig2}. c. The thermal bolometer, using a conventional absorber, has a relatively slow response that is limited by thermalization through heat diffusion and dissipation, which also implies low sensitivity and operation only at low temperature~\cite{liu2016principles}. For example, for a recently proposed graphene nanomechanical bolometer, the thermal response time is about 2.4 $\mu$s at room temperature~\cite{blaikie2019fast} and even it is 1000 times more sensitive than previous graphene-based hot-electron bolometers~\cite{efetov2018fast}. Besides p-n junctions and thermal bolometers, there are other types of infrared detectors based on semiconductor platforms, which are differentiated by the nature of their absorption mechanism. Table I lists their physical mechanisms and their advantages and disadvantages. Fig\ref{fig1} lists some other photodetectors, including impurity absorption, silicon Schottky barriers, quantum wells and quantum dots divides by their different physical mechanisms.

In light of these limitations, a promising avenue for novel photodetection lies in the utilization of the topological diode effect in topological quantum materials arising from their intrinsic quantum wavefunctions~\cite{zhang2021terahertz,ma2019observation, tanaka2022theory}. This method, marked by a nonlinear I-V curve from second order dc current or photocurrent, outperforms previous approaches by eliminating the necessity for bias voltage, covering a wide spectral range from visible down to terahertz frequencies due to the short relaxation time, which allows carriers to respond more quickly to changes in the electric field, and demonstrating ultra-fast and giant responsiveness. The ultra-fast response comes from the efficient generation and collection of photoexcited carriers. Some topological materials possess linear dispersion relation such as the surface state of topological insulators \cite{hedayat2021ultrafast, leppenen2023linear} and the bulk state of Weyl semimetals \cite{yan2017topological}. Due to the high mobility, photoexcited carriers can be quickly collected, resulting in an ultra-fast response. A giant response is created due to their higher quantum efficiency in materials that support intraband transitions. Compared with photodiodes, whose nature is interband transitions, its quantum efficiency will not be larger than 1 as it can only generate one electron-hole pair when absorbing one photon. Therefore its photocurrent is limited. However, for intraband transitions, the quantum efficiency can be larger than 1 because one photon can excite more than one electron when it happens only in the valence band, and this leads to a giant response.

\section{Theory of nonlinear photocurrent}

Berry curvature~($\Omega$) arises in the study of the quantum mechanical behavior of particles in a periodic potential, such as electrons in a crystal lattice. It plays a crucial role in shaping the electronic properties of materials and leads to observable phenomena, such as various Hall effects. Since the Berry curvature, which is defined as the imaginary part of the quantum geometric tensor $\Omega_{\alpha\beta}= \textrm{Im}(-2\sum_{m\neq n}[\braket{\mu_n|i\partial_{k_\alpha}\mu_m}\braket{\mu_m|i\partial_{k_\beta}\mu_n}])$, is odd under time-reversal~($\mathcal{T}$), it is usually necessary to break $\mathcal{T}$ symmetry for non-vanishing Hall current in materials, for example by introducing an external magnetic field ~\cite{hall1880rotational, cage2012quantum, nagaosa2010anomalous, nakatsuji2015large, machida2010time, yasuda2016geometric}. However, in the case of nonlinear photocurrent, a transverse current response can still be obtained even when time-reversal symmetry ($\mathcal{T}$) is present by breaking inversion symmetry~($\mathcal{I}$). Generally speaking, there are two mechanisms for intrinsic nonlinear responses that are directly related to the quantum wavefunctions: the nonlinear Hall effect~\cite{sodemann2015quantum,zhang2018berry,ma2019observation} from intraband transitions~\cite{sato2018role, vardi2009photocurrent, golub2018circular}~(photon energy less than 100 meV below the Drude peak) and the photogalvanic effect from interband transitions~\cite{sipe2000second, de2017quantized, morimoto2016topological, chan2016chiral, ishizuka2016emergent, young2012first, parker2019diagrammatic, xu2018electrically, osterhoudt2019colossal, rees2020helicity, ni2021giant, ma2019nonlinear, ma2017direct, konig2017photogalvanic} (photon energy more than 100 meV).

\subsection{Intraband transition}

\begin{figure*}[pos=!h]
	\centering
		\includegraphics[width=1\textwidth]{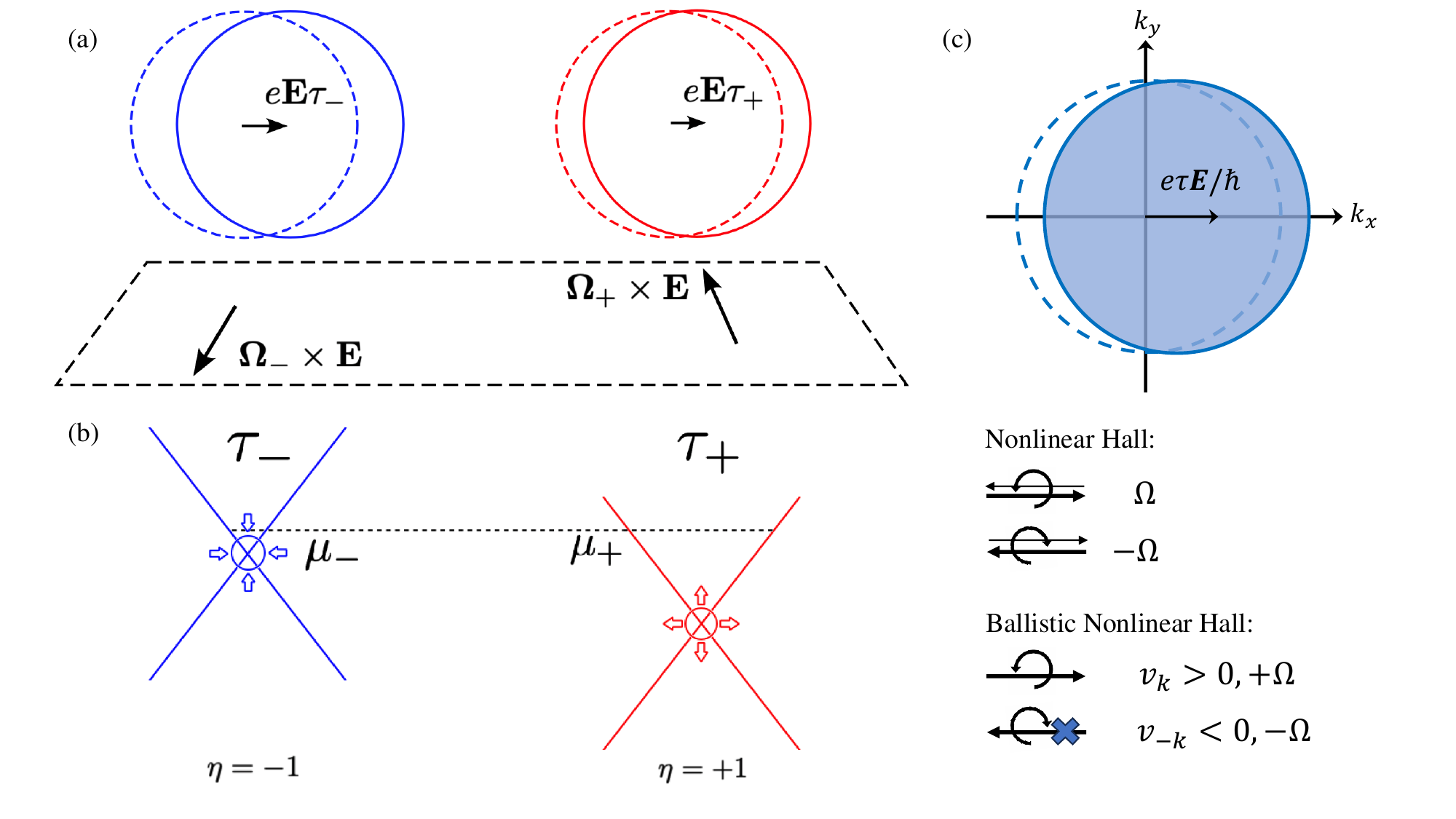}
	  \caption{\textbf{Nonlinear Hall effect in various regimes.} \textbf{a}, The Fermi surface shift demonstrates a direct proportionality to the node-specific momentum relaxation time, elucidating the relationship. Furthermore, the anomalous Berry curvature velocities exhibit divergent directions for nodes of opposite chirality, unveiling contrasting dynamics. \textbf{b}, Berry curvature monopoles result from mirror symmetry breaking. \textbf{c}, Fermi surface shift is unaffected by relaxation time. The Ballistic transport only contains forward modes, without back-scattering. This figure was adapted with permission from \cite{peshcherenko2023chiral}.} \label{fig3}
\end{figure*}

First, we discuss the second-order nonlinear Hall effect~(NHE) from intraband transition, in analogy with conventional Drude current from Fermi surface scattering. 
The NHE is a current-induced anomalous Hall effect and is mainly determined by the Berry curvature dipole~\cite{sodemann2015quantum, low2015topological}. 
The nonlinear response current is given by
\begin{equation}
    \bm{j}^{A}=\frac{e^2}{\hbar}\int\frac{d^{d}\bm{k}}{~(2\pi)^d}f(\bm{k})~(\bm{E}\times\bm{\Omega_k}),
\end{equation}
where $E$ is the external electric field, $f(\bm{k})$ is the full electron distribution function, which depends on the electric field. Under broken inversion symmetry, nonzero Berry curvature can be expected in momentum space. However, due to the existence of $\mathcal{T}$-symmetry, the distribution of Berry curvature obeys $\bm{\Omega_\bm{k}}=-\bm{\Omega_{-\bm{k}}}$~\cite{xiao2010berry}. Therefore, to first order in $\bm{E}$,  there is no net Berry curvature accumulation over the Brillouin zone, leading to zero Hall current response. However, the response current can be nonzero at second order in $\bm{E}$ because there will be nonzero Berry curvature dipole when averaging over a driven Fermi surface (See Fig. \ref{fig3}c). To see this, we can expand the distribution function $f(\bm{k})$ to first order in $\bm{E}$ around the equilibrium distribution $f^0$
\begin{equation}
    f(\bm{k})=f^0(\bm{k})+\frac{e\bm{E}\tau}{\hbar}\cdot~(\partial_{\bm{k}}f^0) + \mathcal{O}(\bm{E}^2),
\end{equation}
where $\tau$ is the scattering rate in Drude transport. The response current is
\begin{equation}
    j^A_{a}=~(\epsilon_{ab\lambda}\frac{e^{3}\tau B_{\lambda c}}{\hbar^2})E_{b}E_{c}=\chi_{abc}E_{b}E_{c},
\end{equation}
where $\chi_{abc}=\epsilon_{ab\lambda}\frac{e^{3}\tau B_{\lambda c}}{\hbar^2}$ is the third-rank nonlinear conductivity tensor and $B_{\lambda c}$ is the Berry curvature dipole and it is given by
\begin{equation}
    B_{\lambda c}=\int\frac{d^{d}\bm{k}}{~(2\pi)^d}f^{0}(\bm{k})\frac{\partial\Omega_{\lambda}}{\partial k_{c}}.
\end{equation}
Here Berry curvature dipole is nonzero as $\frac{\partial\Omega_{\lambda}}{\partial k_c}\neq\frac{\partial\Omega_{\lambda}}{\partial(-k_c)}$. Noncentrosymmetric crystals and surfaces possessing specific groups ($\{ C_{n},C_{nv} \}$, with $n$=1,2,3,4,6), are characterized to have a polar axis, promising $\mathcal{P}$-symmetry breaking. The polar axis also leads to tilted Dirac or Weyl points, where large $B_{\lambda c}$ can be found.~\cite{sodemann2015quantum, zhang2018berry, zhang2018electrically, facio2018strongly, you2018berry, son2019strain, zhou2020highly, tsirkin2018gyrotropic}. Since $B$ is nonzero, the disparity in occupation across the Brillouin zone $f(\bm{k})\neq f(-\bm{k})$ gives rise to a net Berry curvature. This, in turn, leads to a second-order current.
Considering an external electric field $\bm{E}(t)=\bm{E}\cos(\omega t)$ at low frequency $\omega\ll1/\tau$, the transverse current obeys $j^{A}_a~(t)\propto \cos^{2}(\omega t)$, which is the sum of a DC and 2$\omega$ component. Therefore, we get a DC output $j^{0}=\frac{1}{2}j^{A}_a= \frac{1}{2}\chi_{abc} E_b E_c$. This topological Hall rectifier relies only on the intrinsic nonlinearity of the material's response and does not require a finite bias electrical field. Thus it is ideal for passive detection of small signals in terahertz range without power input. Since the incident photon energy is well below the threshold of interband transition, the second-order direct current shows a Drude-like response~\cite{moore2010confinement} as
\begin{equation}
    j^{0}_{a}=~(\frac{1}{1+\omega^{2}\tau^{2}})\frac{\chi_{abc}}{2}E_{b}E_{c},
\end{equation}

\noindent
\textbf{Chiral asymmetry and nonlinear Hall effect from Berry curvature monopole.} Going beyond the constant relaxation time approximation, we obtain the nonlinear Hall current $\mathbf{j}$ in Weyl semimetal as the sum of the contributions from different Weyl pockets (indexed by $\eta$) as \cite{peshcherenko2023chiral}: 
\begin{align}
\mathbf{j}_H(\omega_j)=-e^2\int\frac{d\omega}{2\pi}\sum_\eta\sum_\mathbf{k}\mathbf{\Omega}_\eta\times\mathbf{E}(\omega_j-\omega)\frac{e\mathbf{E}(\omega)\nabla_\mathbf{k}f_0}{-i\omega+1/\tau_\eta},
\label{eq:curr}
\end{align}
where frequency of nonlinear Hall current $\omega_j$ is either dc $\omega_j=0$ or $\omega_j=2\omega$, with $\omega$ is the frequency of the incident the electric field. In Eq. \eqref{eq:curr}, we remove the linear in electrical field $\mathbf{E}$ contribution to Hall current since it is proportional to the net Berry curvature $\sum_\eta\mathbf{\Omega}_\eta$, which vanishes in TR invariant systems. The second order Hall current in Eq. \eqref{eq:curr} arises from anomalous velocity and the shift of Fermi surface under applied electric field $\mathbf{E}$ (see Fig. \ref{fig3}a, b). In the extreme chiral limit with $\tau_{\eta}>>\tau_{-\eta}$, we have a net second order Hall current from Berry curvature monopole, as a chiral version of so-called 'Berry curvature dipole' \cite{sodemann2015quantum,facio2018strongly,zhang2018electrically,you2018berry} nonlinear Hall current. In this scenario, chirality imbalance leads to a node-dependent relaxation time $\tau_\eta$ to promote the contribution of single Berry curvature monopole.

Apart from the Berry curvature monopole and dipole, the nonlinear Hall effect can be induced by other mechanisms such as skew scattering and side jump. For two dimensional materials with specific point groups like $C_{3v}$, a Berry curvature dipole induced in-plane nonlinear Hall effect is forbidden by symmetry. However,  second-order nonlinear current in diffusive systems can still exist due to skew scattering and side jump resulting from the inherent chirality of Bloch electron wavefunctions \cite{isobe2020high}. Further in the clean limit, back scattering as shown in Fig \ref{fig3}c is absent in the ballistic transport. Hence, the nonlinear Hall current is exclusively determined by the integral of Berry curvature over the half of the Fermi surface in the direction of transport \cite{papaj2019magnus}, as illustrated in Fig. \ref{fig3}c for the ballistic non-linear Hall effect. This effect also probes the inherent characteristics of the Bloch electron wavefunctions. Whatever the microscopic origin \cite{du2019disorder,du2021quantum}, a second-order nonlinearity can be used to convert oscillating electric fields into direct current, enabling long-wavelength photodetection below the Drude peak.

\begin{figure*}[pos=!h]
	\centering
		\includegraphics[width=1\textwidth]{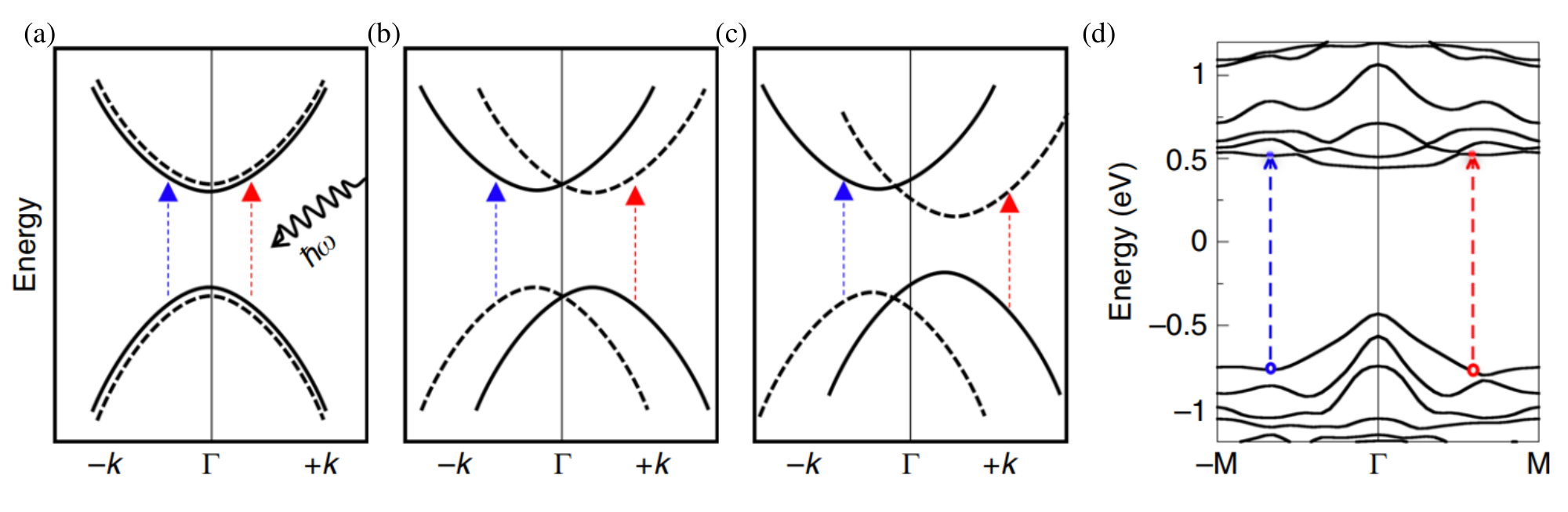}
	  \caption{\textbf{Schematic illustration of the interband transition.} \textbf{a}, 
      In systems exhibiting both inversion symmetry ($\mathcal{P}$) and time-reversal symmetry ($\mathcal{T}$), interband photocurrents are universally prohibited. \textbf{b},In the presence of only $\mathcal{T}$, interband shift current and injection current are enabled. \textbf{c}, In the presence of both $\mathcal{P}$ and $\mathcal{T}$ symmetry breaking, giving rise to a photocurrent, which is distinct from the shift current. \textbf{d}, The band structure of bilayer CrI$_3$ exemplifies the preserved joint $\mathcal{PT}$ symmetry, despite the simultaneous breaking of both $\mathcal{P}$ and $\mathcal{T}$ symmetries. In this case, the violation of the \textbf{k} to -\textbf{k} symmetry is observed. This figure adapted with permission from \cite{zhang2019switchable}.} \label{fig4}
\end{figure*}

\subsection{Interband transition}

In this section, we focus on the photogalvanic effect due to interband transitions. The schematics of interband transition is shown as Fig. \ref{fig4}. The photogalvanic effect consists of two parts: the circular photogalvanic effect~(CPGE)~\cite{de2017quantized, hosur2011circular, ji2019spatially} which is induced by circular polarized light and the shift current effect~\cite{young2012first, tan2016shift, nakamura2017shift} which is induced by linear polarized light. 
The photoconductivity can be written as a general Kubo formula in the velocity gauge~\cite{zhang2018photogalvanic, kraut1979anomalous, von1981theory, kristoffel1980some},
\begin{equation}
    \sigma^{c}_{ab} =\frac{|e|^{3}}{8\pi^{3}\omega^2}\textrm{Re}\{\phi_{ab}\sum_{\Omega=\pm\omega}\sum_{l,m,n}\int_{\textrm{BZ}}d^{3}k~(f_{l}-f_{n}) \times\frac{\bra{n\vec{k}}\hat{v}_a\ket{l\vec{k}}\bra{l\vec{k}}\hat{v}_b\ket{m\vec{k}}\bra{m\vec{k}}\hat{v}_c\ket{n\vec{k}}}{~(E_{n}-E_{m}-i\delta)~(E_{n}-E_{l}+\hbar\Omega-i\delta)}\bm{\}},
\end{equation}
The conductivity is a third rank tensor and the response photocurrent can be written as $\bm{J}^{c}=\sigma^{c}_{ab}E^{*}_{a}E_b$, where $E_a$ and $E_b$ belong to the external electric field $\bm{E}$. In this equation, the velocity term is given by $\hat{v}_a=\frac{\hat{p}}{m_0}, E_n=E_{n}~(\vec{k})$, and further $m_0, \delta=\hbar/\tau$, and $\tau$ stand for the free-electron mass, broadening parameter, and lifetime of a quasiparticle. Here we can make the distinction between circularly polarized light and linearly polarized light by the phase factor $\phi_{ab}$, which represents the phase difference of $E_a$ and $E_b$. $\phi_{ab}=i$ describes the photoconductivity under circular polarized light propagating orthogonally to the plane of $E_a$ and $E_b$. For $\phi_{ab}$ real, it describes the photoconductivity under linearly polarized light.
We analyse the change in this tensor under time-reversal symmetry. Define $N=\bra{n\vec{k}}\hat{v}_a\ket{l\vec{k}}\bra{l\vec{k}}\hat{v}_b\ket{m\vec{k}}\bra{m\vec{k}}\hat{v}_c\ket{n\vec{k}}$. The time-reversal operator $\hat{T}$ reverses all three velocity operators and complex conjugates N. For the tensor to be time-reversal invariant, the real part must vanish, which means $N$ is purely imaginary. When $l=n$ or $n=m$, there is no current, so based on the band number $l$ and $m$, we can partition the contribution into two distinct parts. When $l \neq m$, there are three distinct bands and this is a three-band process~($n\rightarrow m\rightarrow l$). When $l=m$, this is a two-band process. 

The conductivity tensor is both related to $\phi_{ab}$ and the integral. We separate the integral into $N$ and the denominator to solve it. The denominator can be written as $D_1 D_2$, with
\begin{equation}
\begin{split}
    & D_{1}=\frac{1}{E_{n}-E_{m}-i\delta}=\frac{P}{E_{n}-E_{m}}+i\pi\delta~(E_{n}-E_{m}), \\
    & D_{2}= \frac{1}{E_{n}-E_{l}+\hbar\Omega-i\delta}=\frac{P}{E_{n}-E_{l}+\hbar\Omega}+i\pi\delta~(E_{n}-E_{l}+\hbar\Omega).
\end{split}
\end{equation}
Where $P$ denotes the Cauchy principal value. First, we consider $\phi$ to be real, the linearly polarized case. As the conductivity tensor only contains the real part, in this case, we should take the imaginary part (since N is imaginary) $\textrm{Im}(D_{1}D_{2})\sim \pi\frac{P}{E_{n}-E_{m}}\delta~(E_{n}-E_{l}+\hbar\Omega)$, enforcing  that only momentum vectors with a band gap equal to the photon energy~($|E_{n}-E_{l}|=\hbar\omega$) contribute to the shift current effect. Therefore, the shift current also contributes to response current only in some small, select regions in momentum space. Next we consider $\phi=i$, the circularly polarized case. In this process, real part of the denominator should be taken into account, $\textrm{Re}(D_{1}D_{2}) \sim \frac{1}{~(E_{n}-E_{m})~(E_{n}-E_{l}+\hbar\Omega)}$.

Further, the three band equation~(6) can be simplified to the form of interband Berry curvature for CPGE~\cite{sipe2000second, de2017quantized,de2020difference}, and here it is represented as the rate of current generation,
\begin{equation}
    \frac{dj_a}{dt}=\beta_{ab}~(\omega)\left[\bm{E}~(\omega)\times\bm{E}^{*}~(\omega)\right]_b,
\end{equation}
where $\beta_{ab}$ is the circular photogalvanic conductivity tensor:
\begin{equation}
    \beta_{ab}(\omega)=\frac{i\pi e^3}{4\hbar}\int_{\textrm{BZ}}\frac{d\bm{k}}{(2\pi)^2}\sum_{n>m}\epsilon^{bcd}f_{nm} \times\Delta_{mn}^{a}\textrm{Im}[r_{nm}^{d}r_{mn}^c]\mathcal{L}_{\tau}~(E_{nm}-\hbar\omega).
\end{equation}
In this equation, $E_{mn}=E_{m}-E_n$ represents the difference of the $n$ and $m$ band dispersion relation, $f_{mn}=f_{m}-f_n$ represents the difference in band distribution, $\Delta_{mn}^{a}=\partial_{k_a}E_{mn}/\hbar$ is the difference in group velocities, and $r_{mn}^{a}=i\braket{m|\partial_{k_a}|n}$ is the off-diagonal Berry connection. To incorporate a finite relaxation time $\tau$, we utilize the Lorentzian function $\mathcal{L}_{\tau}(E_{nm}-\hbar\omega)$, where $\mathcal{L}_{\tau} = \frac{1}{\pi}\frac{\eta}{\omega^{2}+\eta^2}$ and $\tau = \hbar/\eta$. This representation aligns with the form of Berry curvature, which can be expressed as
\begin{equation} 
    \bm{\Omega_{n}^{b}}=\epsilon^{bcd}\sum_{n>m}\textrm{Im}\left[r_{nm}^{d}r_{mn}^c\right].
\end{equation}
Equation~(6) can be also written similarly as the form of Berry connection for shift current under linear polarized light~\cite{young2012first,de2020difference}
\begin{equation}
    \sigma_{abd}(E)=\frac{\pi C}{2}\int_{k}\sum_{n>m}f_{nm}\textrm{Im}\left[r_{mn}^{b}r_{nm;a}^{c}\right]\delta~(E_{nm}-E),
\label{eq:pc}
\end{equation}
where $C=e^{3}/\hbar^2$, $\int_{k}=\int d^{3}k/~(2\pi)^3$, the Bloch eigenstates are $H\ket{n}=E_{n}\ket{n}$, and $r_{nm;a}^{b}=\partial_{k_a}r_{nm}^{b}-i~(\xi_{nn}^{a}-\xi_{mm}^{a})r_{nm}^b$ with $\xi_{nn}^{a}=i\braket{n|\partial_{k_a}|n}$ as the diagonal Berry connection. Equation~(6) shows that photoconductivity is also related to the charge distribution and Berry curvature. Thus, only inversion symmetry breaking is required and the requirements on point groups are less strict compared to nonlinear Hall effects of Berry curvature origin.

\begin{figure*}[pos=!h]
	\centering
		\includegraphics[width=1\textwidth]{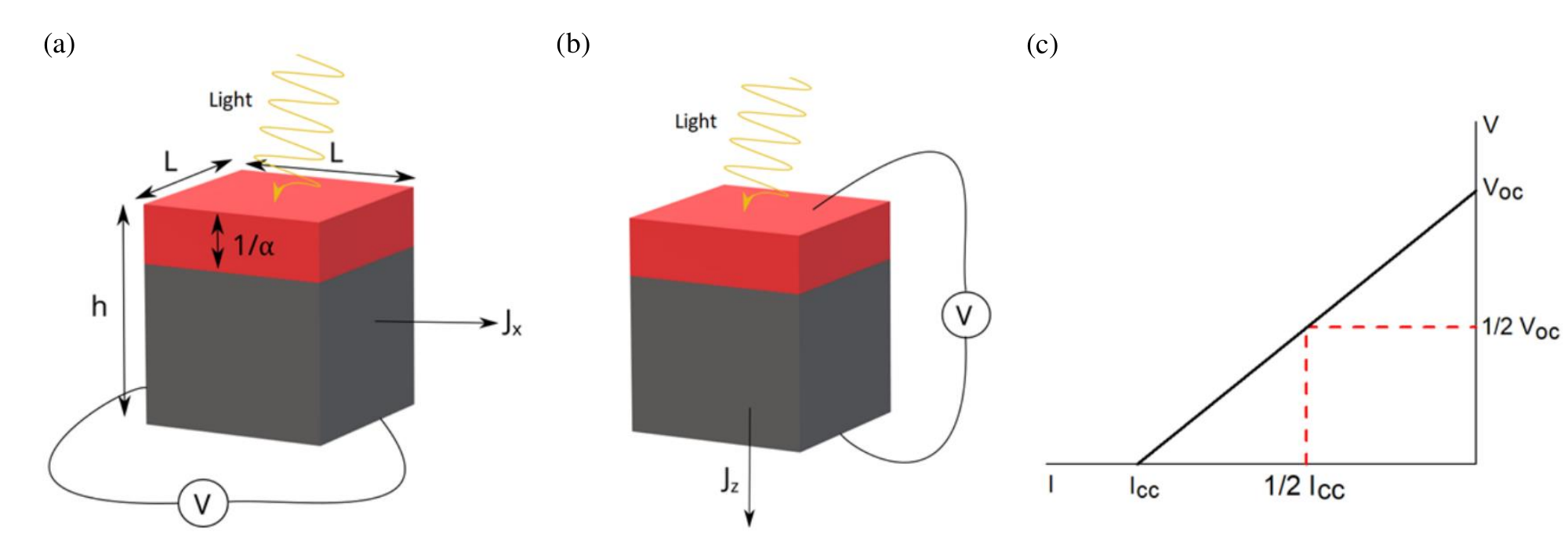}
	  \caption{\textbf{Schematic photodevice and its I-V diagram.} \textbf{a}, \textbf{b}, Sample structure and two possible current geometries. \textbf{c}, Voltage (V) and current (I). They are linear with each other here. The red dashed lines denote the values that optimize energy harvesting efficiency. This figure adapted with permission from \cite{campos2022intrinsic}.}\label{fig5}
\end{figure*}

\subsection{Energy conversion and current responsivity}
With the theory of the dc response for the nonlinear Hall current and interband photocurrent, we now delve into the physical principles governing the energy conversion efficiency and current 
responsivity as the figures of merit for real devices, before proceeding to review experimental findings. For a realistic photodevice with thickness $h$ and lateral dimensions $L$, as shown in Fig. \ref{fig5}, its LPGE response to light can be written as~\cite{campos2022intrinsic, tan2016shift, morimoto2018current} 
\begin{equation}
    J_i=J_i^{~(2)}=\frac{k_i}{\alpha h}\mathcal{I}_0,
\end{equation}
where $k_i=\frac{4}{c\epsilon_0}\sigma_{ijk}^{(2)}\widetilde{E}_{j}\widetilde{E}_{k}$, $\sigma_{ijk}^{(2)}$ is the third rank photoconductivity given by equation \ref{eq:pc}, $\widetilde{E}_j$ gives a polarization, $\mathcal{I}_0$ is the maximum light intensity and $\alpha$ is the attenuation coefficient. Here $h\gg 1/\alpha$ for a bulk sample. For a given direction the corresponding open circuit voltage is $V_{oc}=E_i^0d$, where $E_i^0=k_i / \Sigma_{ii}$ depends on the fourth order conductivity tensor as
\begin{equation}
\Sigma_{ii}=\frac{4}{c\epsilon_0}\sigma_{ijki}^{(3)} E_j E_k =-\tau^2\frac{4}{c\epsilon_0}\frac{\pi e^4}{18\hbar^3}\int \frac{d^3k}{(2\pi)^3}\sum_{n\neq m}f_{mn}\partial_i\partial_i\epsilon_{nm}Re[r^i_{nm}r^k_{nm}]\delta(\hbar\Omega-\epsilon_{nm}),
\end{equation}
and $d$ is the distance across which the voltage drop occurs. The total power converted by the device is $P=IV$ and the power of electromagnetic field of light is $W=\mathcal{I}_0L^2$, so the energy conversion efficiency can be defined as $\eta=P/W$. As shown in Fig. \ref{fig5}, $I$ is linear in $V$ for LPGE, so the total power $P$ reaches maximum when $I=\frac{1}{2}I_{cc}$ and $V=\frac{1}{2}V_{oc}$, giving $P=\frac{1}{4}I_{cc}V_{oc}$. Thus, the energy conversion efficiency is
\begin{equation}
    \eta=\frac{1}{4}\frac{~(J_i^{~(2)}A)~(E_i^0d)}{\mathcal{I}_0L^2}=\frac{1}{4}\frac{k_i^2}{\Sigma_{ii}\alpha h}\frac{Ad}{L^2}=\frac{1}{4}\frac{k_i^2}{\Sigma_{ii}\alpha},
\end{equation}
In shift current, where $k_i$ remains independent of carrier lifetime $\tau$, and $\Sigma_{ii}$ scales as $\propto\tau^2$, the efficiency diminishes as $1/\tau^2$ in the limit. Thus, only the injection current survives, representing ballistic currents in the relaxation time approximation at zero temperature~\cite{sipe2000second, aversa1995nonlinear, nastos2010optical}. In the case of injection current, where $k_i$ scales linearly with $\tau$ and $\Sigma_{ii}$ scales quadratically with $\tau$, its energy conversion efficiency remains unaffected by $\tau$.

For the purpose of photodetection, a well-accepted figure of merit is the current or voltage responsivity. Here we discuss the case of nonlinear Hall current from Berry curvature dipole for simplicity. In a homogeneous crystal material, the current responsivity is defined by comparing the response dc nonlinear Hall current with the power absorbed by the nonlinear Hall rectifier as:
\begin{equation}
    R=\frac{j^0}{W\sigma_{ab}E_{a}E_{b}},
\end{equation}
where $W$ is the width of the Hall rectifier and $\sigma_{ab}$ is the general nonlinear Hall conductivity,
\begin{equation}
    \sigma_{ab}=\frac{e^{2}D_{ab}}{\hbar^2}\frac{\tau}{~(1+\omega^{2}\tau^2)},
\end{equation}
with
\begin{equation}
    D_{ab}=\int\frac{d^{d}k}{~(2\pi)^d}\frac{\partial E}{\partial k_a}\frac{\partial E}{\partial k_b}\left(-\frac{\partial f_0}{\partial E}\right),
\end{equation}
$R$ is independent of incident power, frequency, and scattering rate, as $j^0\propto\frac{\tau}{1+\omega^2\tau^2}$, only depending on the Berry curvature dipole $B$ and Drude weight $D$, which means $R$ can be predicted just by intrinsic properties of material. Considering sensing range, temperature and responsivity, different types of materials have different performance. In that case, it is necessary to analyse different types of materials for better efficiency in photodetection. 

\begin{figure*}[pos=!h]
	\centering
		\includegraphics[width=0.9\textwidth]{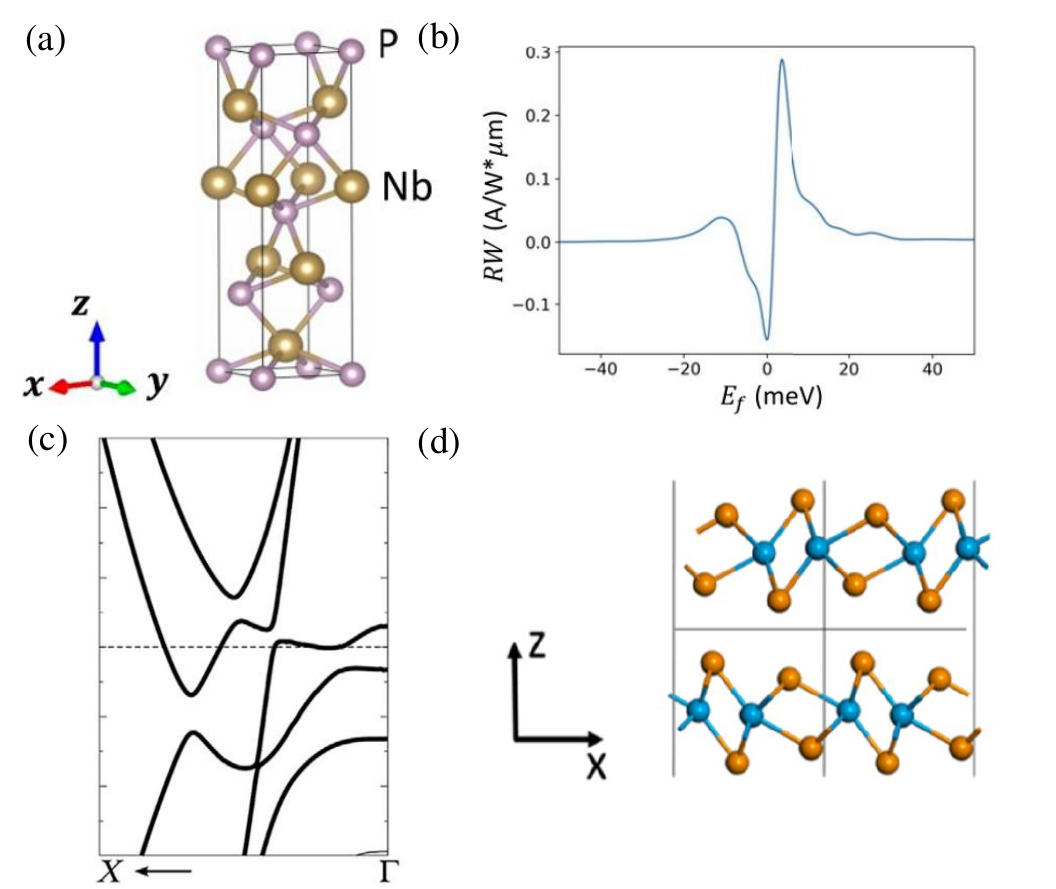}
	  \caption{\textbf{Theory of Weyl semimetal.} \textbf{a}, Lattice structure of NbP. \textbf{b}, Responsivity of current with respect to chemical potential of NbP. \textbf{c}, Band structure of bilayer WTe$_2$. \textbf{d}, Lattice structure of WTe$_2$. Panels \textbf{a} and \textbf{b} adapted with permission from \cite{zhang2021terahertz}, panel \textbf{c} adapted with permission from \cite{muechler2016topological} and panel \textbf{d} adapted with permission from \cite{yang2018origin}.}\label{fig6}
\end{figure*}

Now we come to the material part. First, we introduce Weyl semimetals~\cite{vazifeh2013electromagnetic, goswami2015optical, kargarian2015theory, ishizuka2016emergent, ishizuka2017momentum, hosur2011circular, sodemann2015quantum, chan2016chiral, morimoto2016semiclassical, taguchi2016photovoltaic, de2017quantized, chan2017photocurrents, konig2017photogalvanic, rostami2018nonlinear, golub2017photocurrent, zhang2018berry}, specially highlighting NbP~\cite{zhang2021terahertz, min2023colossal} as an example. Nonlinear photocurrent, as discussed earlier, necessitates the breaking of inversion-symmetry. Therefore, Weyl semimetals like TaAs, TaP, NbAs and NbP~\cite{bansil2016colloquium, hasan2017discovery, weng2015weyl, armitage2018weyl}, lacking inversion symmetry and hosting Weyl nodes with divergent Berry curvature, are naturally suitable materials. NbP possesses the point group symmetry $C_{4v}$ with a rotation axis $z$ and two mirror plane $xz$ and $yz$, as depicted in Fig. \ref{fig6}a. Therefore, the response current of NbP attains its maximum when the external electric field lies in the $xy$ plane, and the measured current is in the $z$ direction, with Berry curvature dipole tensor $B_{yx}=-B_{xy}$ are nonzero.
In NbP, the theoretically calculated responsivity $R$ is
\begin{equation}
    R=\frac{eB_{xy}}{2WD_{xx}},
\end{equation}
In Fig. \ref{fig6}b, a notable responsivity is evident within $\pm$20 meV around the Fermi energy range, particularly near two extended loops in momentum space with large Berry curvature dipole.

In addition to Weyl semimetals with point group $C_{4v}$, other Weyl semimetals with point group symmetry $C_{2v}$, like MoTe$_{2}$ and WTe$_{2}$, ~(as depicted in Fig. \ref{fig6}c,d)~\cite{soluyanov2015type, ortix2021nonlinear, muechler2016topological, yang2018origin}, can also host nonlinear Hall current. However, due to the reduced Drude weight arising from small electron (hole) pockets, Weyl semimetals with point group $C_{4v}$ usually exhibit larger responsivity.

\begin{figure*}[pos=!h]
	\centering
		\includegraphics[width=0.9\textwidth]{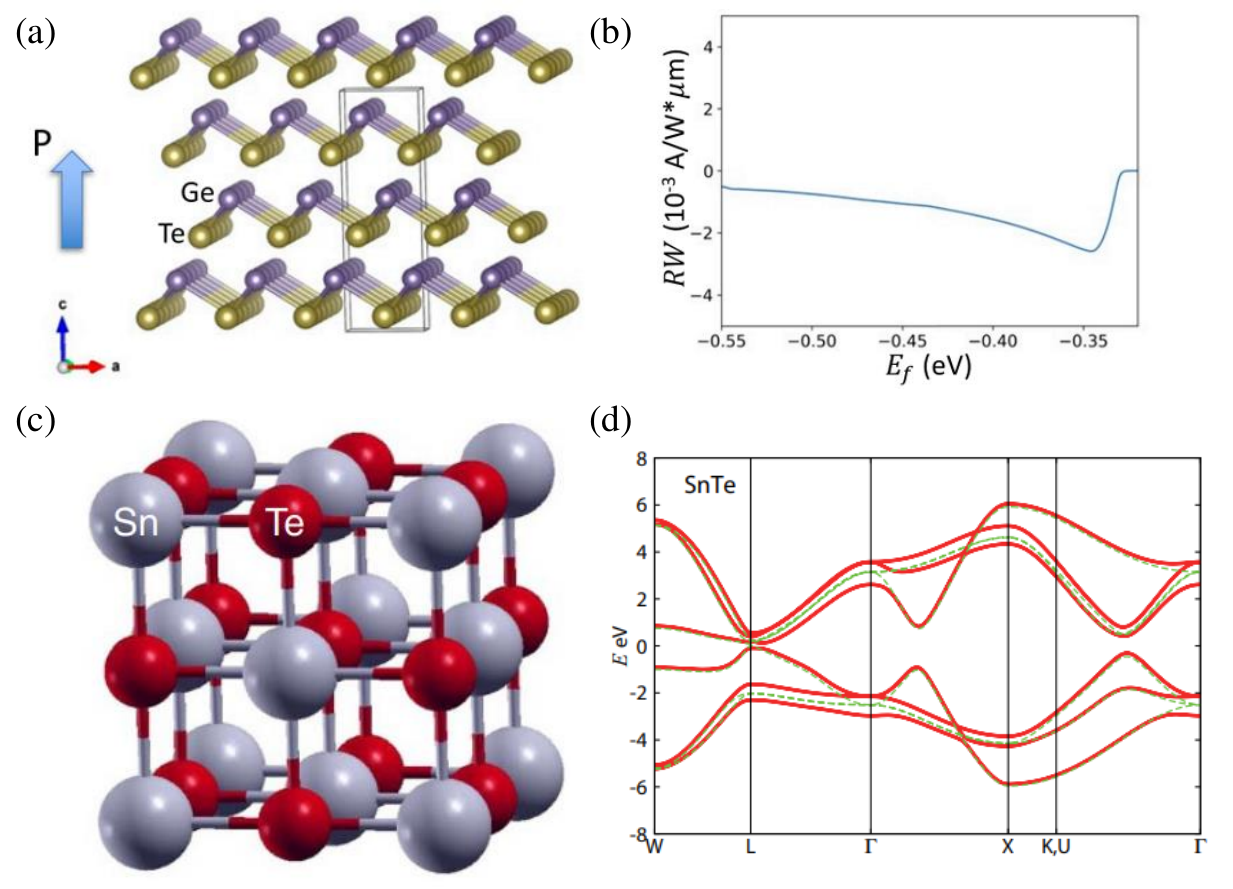}
	  \caption{\textbf{Theory of ferroelectric topological materials.} \textbf{a}, Crystal structure of GeTe. The polarization is along c axis. \textbf{b}, Responsivity of GeTe. \textbf{c}, Crystal structure of SnTe. \textbf{d}, Band structure of SnTe, where solid lines represent results from first-principles calculations, while dotted lines depict those derived from the tight-binding model. The left bottom part is the Brillouin zone. Panels \textbf{a} and \textbf{b} adapted with permission from \cite{zhang2021terahertz}, panel \textbf{c} adapted with permission from \cite{hsieh2012topological} and panel \textbf{d} adapted with permission from \cite{kuraya2015band}.} \label{fig7}
\end{figure*}

Next, we introduce topological ferroelectric materials~\cite{zhang2021terahertz}. The ferroelectric monochalcogenides GeTe, and SnTe exhibit a rhombohedrally distorted rock salt structure, which belong to R3m, a kind of group $C_{3v}$. GeTe, illustrated in Fig. \ref{fig7}a, boasts a high Curie temperature up to 700 K and a global bandgap that reaches 0.32 eV. With a Berry curvature dipole $B_{xy}$ reaching 0.5, GeTe exhibits substantial current responsivity across a wide range of dopings. At photon energies exceeding the Drude width ~(40 meV), the DC photocurrent is predominantly of interband origin as shift and injection currents. Based on first-principles calculation, the photocurrent responsivity from the mid infrared to near infrared range reaches values in excess of 3 $\times$ 10$^{-4}$ $\mu$mA/W. This value is an order of magnitude larger than conventional ferroelectric semiconductors like BaTiO$_{3}$ and SbSI~\cite{sotome2019spectral}. The interband responsivity of GeTe is shown in Fig. \ref{fig7}b. Moving to another topological ferroelectric insulator, SnTe, whose lattice structure and band structure are presented in Fig. \ref{fig7}c and Fig. \ref{fig7}d, respectively, exhibits a narrow bandgap, tilted Weyl points, and multiple surface states~\cite{fu2011topological}, making it highly efficient for photodetection in the near and IR spectral range. Moreover, the bandgap of GeTe and SnTe can be tuned by adjusting Ge and Sn concentrations. This interplay between ferroelectric distortion, chemical composition, and topological phase transition in Ge$_{1-x}$Sn$_{x}$Te may significantly impact its photocurrent response in the THz and far-infrared frequency ranges.

\begin{figure*}[pos=!h]
	\centering
		\includegraphics[width=0.9\textwidth]{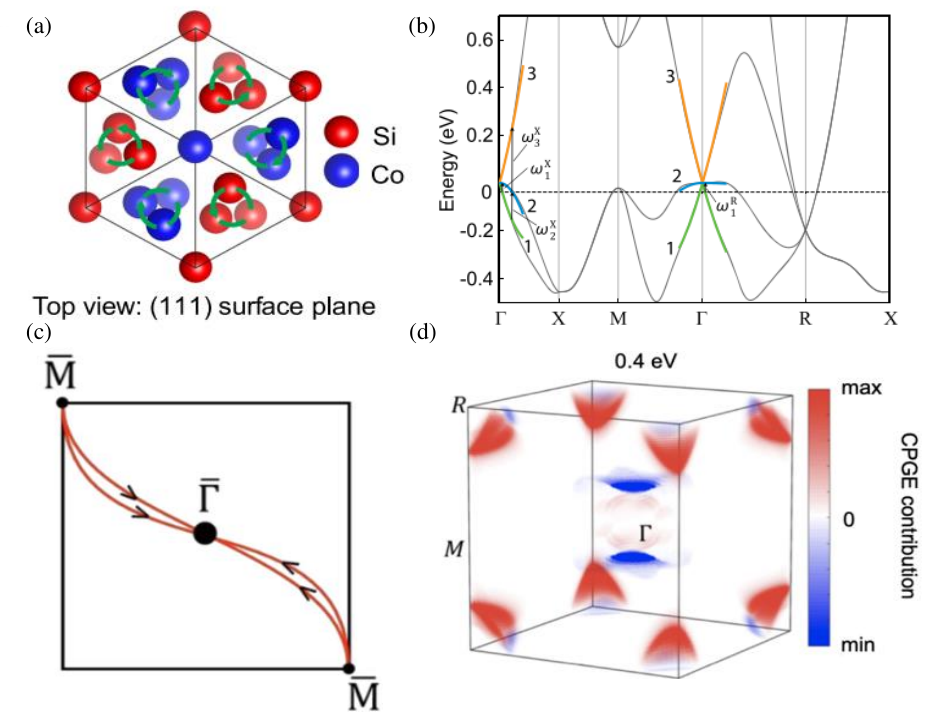}
	  \caption{\textbf{Theory of topological chiral crystal.} \textbf{a}, Schematic top-down representation of the CoSi (111) crystal surface. \textbf{b}, 
      The band structure of CoSi in the absence of spin-orbit coupling, with zero energy positioned at the threefold node located at the $\Gamma$ point. \textbf{c}, Schematic figure for long surface Fermi arc in CoSi. \textbf{d}, Momentum-resolved contributions to the CPGE peak when the photon energy is 0.4 eV, revealing the activation of two Weyl nodes. Panels \textbf{a}, \textbf{b} and \textbf{d} adapted with permission from \cite{ni2021giant} and panel \textbf{c} adapted with permission from \cite{kumar2020topological}.} \label{fig8}
\end{figure*}

Another family of theoretically proposed candidate materials are topological chiral crystals, characterized by linear dispersive bands and multi-fold nodal crossings at the Fermi level, paving the way for a colossal long-wavelength photogalvanic effect. Notably, the longitudinal circular photogalvanic effect (CPGE) in these crystals, especially observed in materials like RhSi and CoSi~(Fig. \ref{fig8}a, b)~\cite{de2017quantized, rees2020helicity,ni2021giant}, is remarkably large due to its quantized nature from chiral topological nodes. 
Near the Fermi surface, CoSi and RhSi have protected nodal crossings, and these nodes are located at different energies when their chiralities are opposite, as all mirror symmetries are broken. Consequently, only the nodal crossing near the Fermi level is activated by terahertz or infrared radiation, while the one with opposite chirality is Pauli-blocked, resulting in a substantial net topological photocurrent. Taking CoSi as an example, the momentum-resolved calculation~(Fig. \ref{fig8}d) reveals that the CPGE peak at 0.4 eV originates solely from the interband transition of its multifold fermions at $\Gamma$ and $R$, exhibiting a clear topological origin. In addition to the interband response, intraband transitions exist for THz and IR frequencies. In the bulk, the Weyl nodes~(Berry curvature monopole) induce a nonzero Berry curvature in the Fermi surface, and these divergent Berry curvature monopoles contribute to a large photocurrent around the Drude width, up to 100 meV. Furthermore, topological chiral crystals hold a long surface Fermi arc, which can serve as another source of optical absorption. Taking CoSi as an example~(Fig. \ref{fig8}c), the Fermi arc significantly influences the linear optical conductivity, and may lead to large surface photocurrent \cite{ni2021giant}. 

\subsection{High-throughput search of inversion breaking topological materials}

The identification of topological states in inversion breaking topological materials poses a challenge because the degenerate points are no longer located at high symmetry points, and may be anywhere in the Brillouin zone. This complexity makes it difficult to search for or predict novel topological materials, especially Weyl semimetals for broadband photodetection. Therefore, a new and effective algorithm to find Weyl points and establish a database is needed. To address this issue, a high-throughput calculation workflow has been developed to search for Weyl semimetals and calculate their nonlinear optical response~\cite{xu2020comprehensive}. This approach starts with crystal structures available in the Inorganic Crystal Structure Database~(ICSD)\cite{hellenbrandt2004inorganic}, for which computational screening was performed to identify nonmagnetic Weyl semimetals with Weyl points near the Fermi level (at about 300 meV). 
  
\begin{figure*}[pos=!h]
	\centering
		\includegraphics[width=1\textwidth]{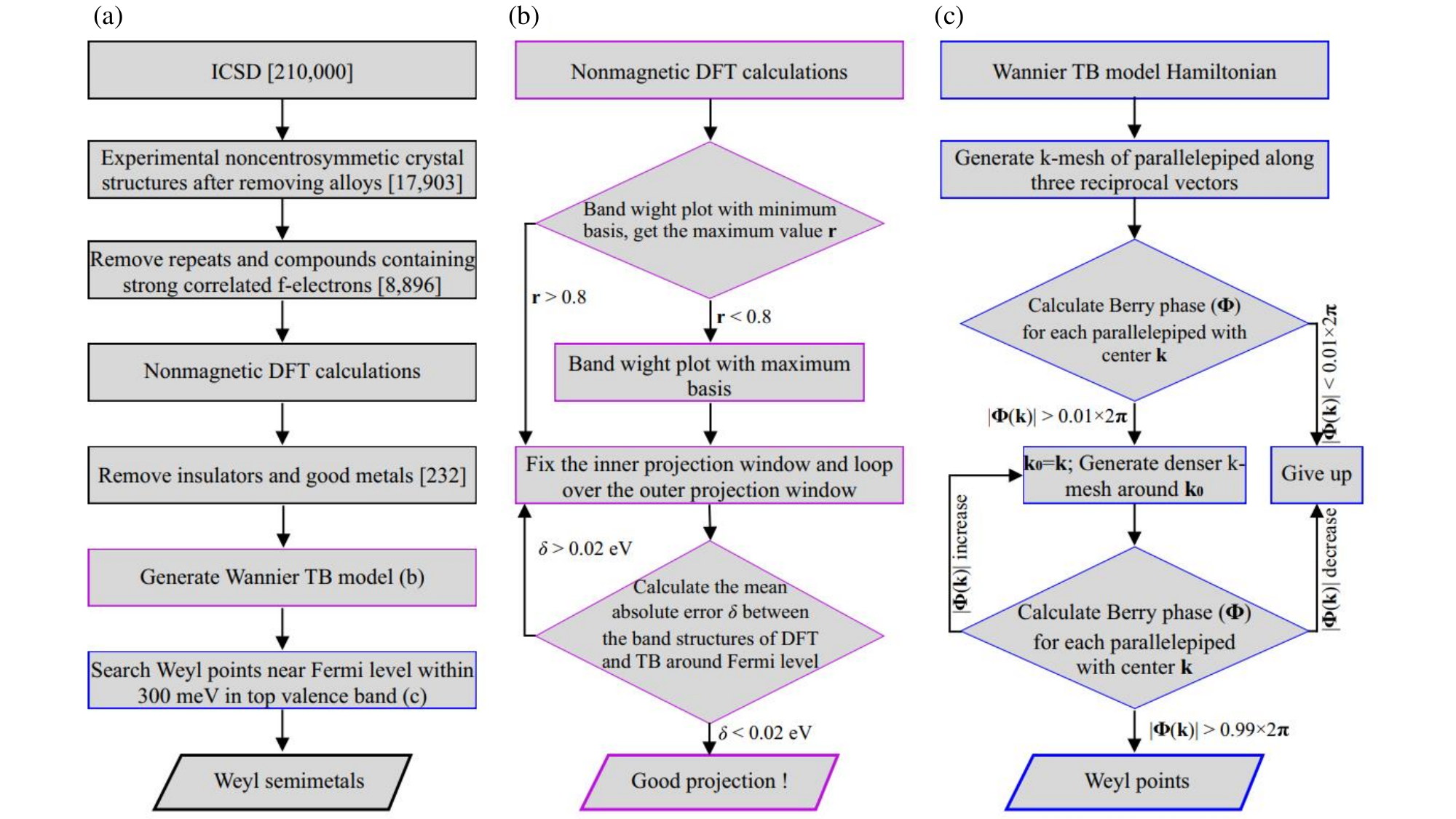}
	  \caption{\textbf{Workflow of the calculations for the Weyl semimetals search and identification.} \textbf{a}, Obtain noncentrosymmetric crystal structures that are proved by experiments from the Inorganic Crystal Structure Database (ICSD), excluding alloys and duplicates. Compounds with strong f-electron correlations are omitted. Bandgap and carrier densities are calculated nonmagnetically. A material is labeled magnetic if it contains magnetic elements or exhibits magnetic moments over 0.05 $\mu$B per unit cell. Among the 232 semimetal phases, Bloch wave functions are transformed into Wannier functions to create tight-binding (TB) model Hamiltonians, which are then used to locate Weyl points near the Fermi energy. \textbf{b}, Automated Wannier function generation branch. \textbf{c}, Utilizing the Berry phase approach, automated Sub-Branch for Weyl points search. This figure adapted with permission from \cite{xu2020comprehensive}.} \label{fig9}
\end{figure*}

The systematic approach to identifying nonmagnetic Weyl semimetals, depicted in Fig. \ref{fig9}, hinges on two core algorithms: automatic Wannier function generation and Weyl point search. The first step of this process is obtaining the experimental noncentrosymmetric crystal structures from the ICSD database and the following steps are based on them. To maintain the reliability of the analysis, systems featuring strongly correlated f-electrons are excluded as density functional theory (DFT) is not very accurate. However systems with 3d-electrons, regardless of their correlation strength, are included to ensure that potential candidates are not overlooked. After filtering out alloys and redundant structures, the workflow retains 8896 inversion-breaking compounds, ready for further investigation.

\begin{figure*}[pos=!h]
	\centering
		\includegraphics[width=1\textwidth]{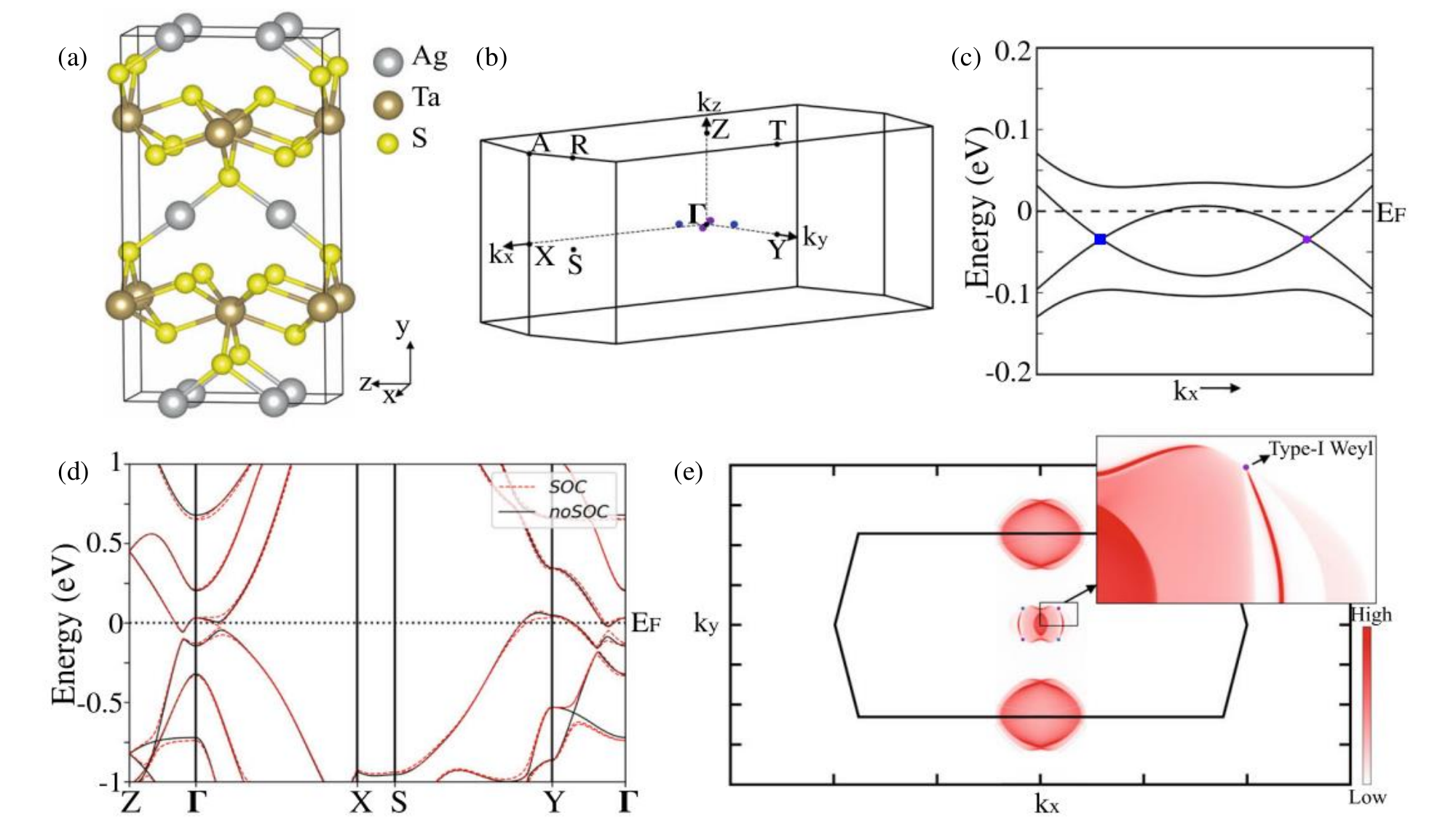}
	  \caption{\textbf{Calculated Weyl semimetals TaAgS$_3$.} \textbf{a}, The crystal structure of TaAgS$_3$. \textbf{b}, By Weyl point search algorithm, the Brillouin zone and the distribution of TaAgS$_3$ can be obtained. \textbf{c}, Energy dispersion along a Weyl point pair aligned parallel to the $k_x$ axis. \textbf{d}, Band structure comparison along high-symmetry lines, showcasing the effect of spin-orbit coupling: red dotted lines represent results with spin-orbit coupling, while black solid lines depict those without. Notably, band inversion occurs near the $\Gamma$ point. \textbf{e}, Surface Fermi arcs correspond to fixed energy levels at Weyl points. Purple dots indicate positive chirality while blue dots indicate negative chirality. This figure adapted with permission from \cite{xu2020comprehensive}.} \label{fig10}
\end{figure*}

For each inversion-breaking compound, the Wannier functions are constructed, as shown in Fig. \ref{fig10}b. Using the highly symmetric tight-binding Hamiltonians derived from the projected Wannier functions, the algorithm scans for Weyl points. A raw list of Weyl semimetals, encompassing all Weyl point information, is obtained, and after thorough processing and double-checking, the list is finalized. In conclusion, 46 nonmagnetic Weyl semimetals with Weyl points within 300 meV of the Fermi level are identified, all classifiable as either type-I or type-II Weyl semimetals.

Out of the computationally identified materials, we specifically mention TaAgS$_3$ as a typical candidate, representing a type-I Weyl semimetal with space group Cmc2$_1$. The lattice structure of TaAgS$_3$ is illustrated in Fig. \ref{fig10}a, c, and its band structure, is shown in Fig. \ref{fig10}d. The band structure unveils a semimetallic configuration characterized by small electron and hole pockets, which locate near the $\Gamma$ and $Y$ points.
 
The Weyl points in TaAgS$_3$ are confined to the $k_z$ plane due to the presence of $\mathcal{T}$ symmetry and C$_{2z}$ group point symmetry, as illustrated in Fig. \ref{fig10}b. Remarkably, this material hosts two pairs of Weyl points, all at the same energy level, roughly 30 meV below the charge-neutral point. When the chemical potential aligns with these Weyl points, Fermi arcs emerge, as shown in Figure \ref{fig10}e. Notably, these Fermi arcs cover about 15\% of the reciprocal lattice vector, making them detectable via surface-sensitive techniques like angle-resolved photoemission spectroscopy (ARPES) and scanning tunneling microscopy (STM).

\subsection{Review of experimental progress}
In addition to a wealth of theoretical predictions, a variety of experimental findings solidify the significant potential of various materials for broadband photodetection. Here, we highlight four main categories of materials as examples: two dimensional topological insulators, Weyl semimetals, chiral crystals, and ferroelectric topological materials.

\subsubsection{Experimental setup}

\begin{figure*}[pos=!h]
	\centering
		\includegraphics[width=1\textwidth]{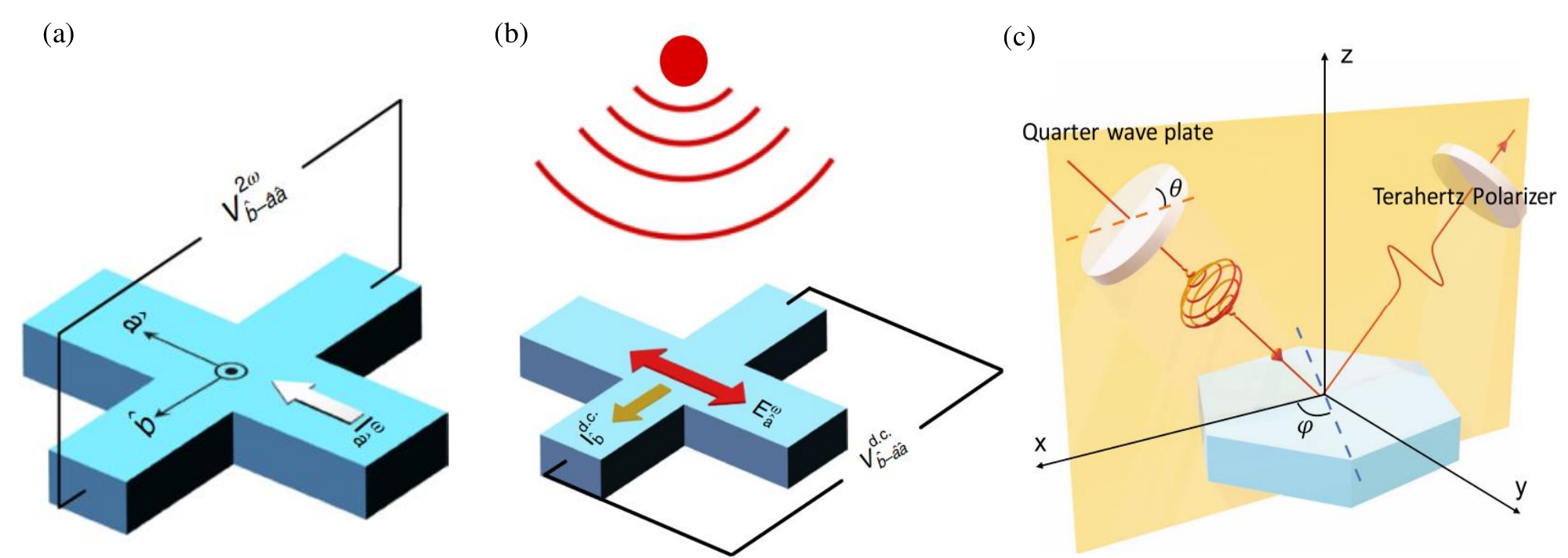}
	  \caption{\textbf{Experiment setup for detecting nonlinear currents and photodetection} \textbf{a}, Structure of Hall bar device and the way to measurement. \textbf{b}, Diagram illustrating the rectifier function. The Hall bar device rectifies the incident electric field along the $\hat{a}$ direction ($E^{\omega}{\hat{a}}$), resulting in a d.c. current along the $\hat{b}$ direction ($I^{d.c.}{\hat{b}}$). \textbf{c}, Schematic illustrating CPGE detection via THz pulse measurements. Panel \textbf{a} adapted with permission from \cite{kumar2021room} and panels \textbf{b} and \textbf{c} adapted with permission from \cite{ni2021giant}.} \label{fig11}
\end{figure*}

The most widely used method for photodetection involves measuring the DC current or voltage response. In differentiating between the nonlinear Hall effect and the photogalvanic effect, the key distinction lies in the polarization requirement. The nonlinear Hall effect does not necessitate specific polarization, while the photogalvanic effect does. Consequently, a photogalvanic effect experiment involves incorporating an additional polarizer into the experimental setup, as depicted in Fig. \ref{fig11}a, which illustrates the measurement configuration and the Hall bar device.

Typically, to assess the presence of nonlinear photocurrent in a material, an AC current is applied to the Hall bar. Once the existence of nonlinear photocurrent is confirmed, it is employed for photodetection experiments, as illustrated in Fig. \ref{fig11}b, using the nonlinear Hall effect as an example. The Hall bar rectifies the incident electric field due to the radiation, generating a DC current along the transverse direction of the electric field. For photogalvanic experiments, a polarizer is introduced to produce circular or linearly polarized light before it is directed onto the rectifier. This additional component is essential for generating the specific polarization required for the photogalvanic effect.

Besides DC current measurement, another way to detect CPGE is detect the radiated THz pulses emitted from illuminated regions, as shown in Fig. \ref{fig11}c. In this setup, a current along the propagation direction is induced by circularly polarized light inside the material. The angle of incidence for the laser pulse is 45 degrees, creating a time-varying surface current and emitting terahertz waves~\cite{rees2020helicity, shan2004terahertz}. An achromatic quarter-wave plate is used here to adjust the polarization of incident light. The terahertz polarizer separates the THz wave into $xz$ and $y$ directions before reaching the detector. The photocurrent, expressed by $j(\omega)=\frac{\beta_{xx}}{i\omega+1/\tau}$, where $\omega$ is the THz frequency, can be obtained from this setup.

\subsubsection{Experimental measurement}

\begin{figure*}[pos=!h]
	\centering
		\includegraphics[width=0.9\textwidth]{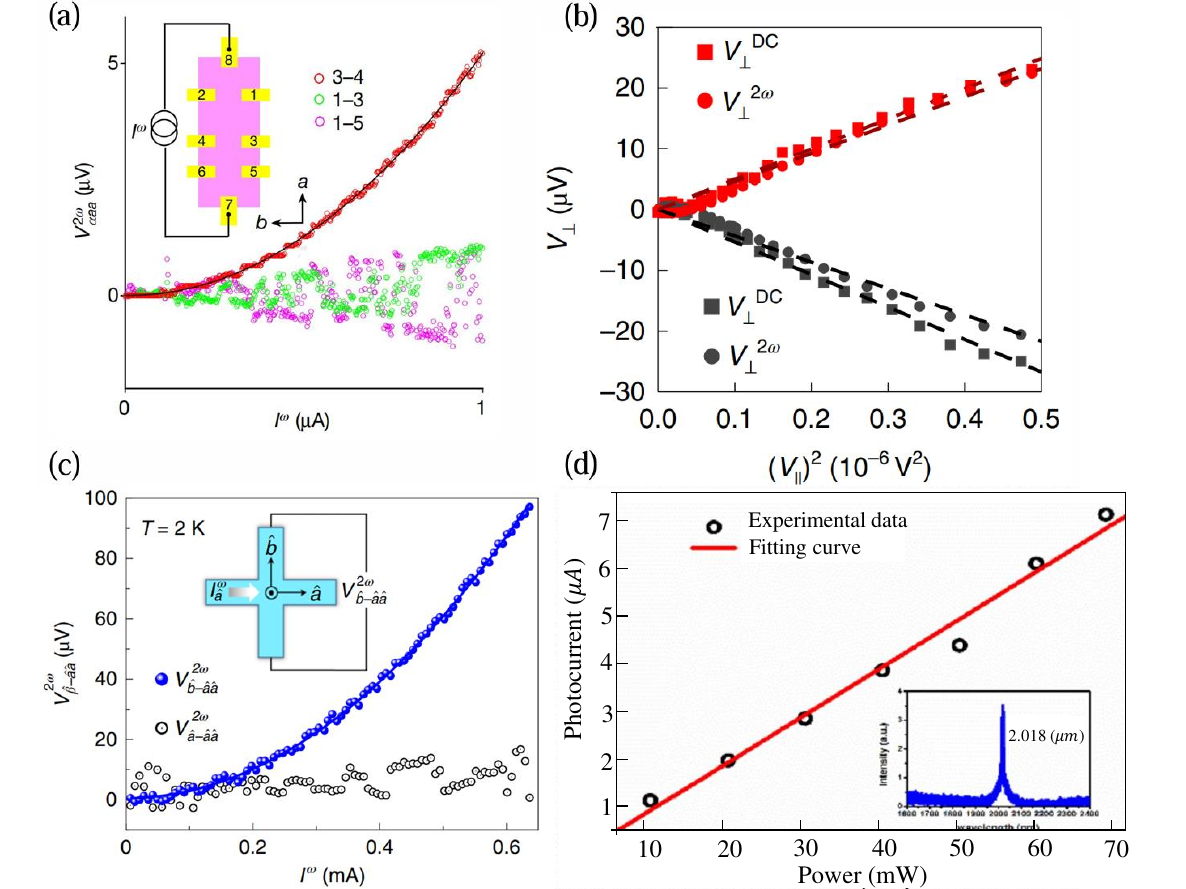}
	  \caption{\textbf{Nonlinear currents in Weyl Semimetals} \textbf{a}, Second-harmonic longitudinal and transverse voltages measured between electrodes \boldmath{7} and \boldmath{8}, in response to an AC current applied along the a direction of bilayer WTe$_{2}$. \textbf{b}, $V^{2\omega}{\perp}$ and $V^{DC}{\perp}$ plotted respect to $V_{\parallel}$ in a tetralayer WTe$_2$ device. The data for forward (red) and backward (black) current are obtained in the generally same environment. The uncertainties are negligible here. Dashed lines represent the linear fits to the experimental result. \textbf{c}, Second-harmonic longitudinal and transverse voltages induced by an AC current along the \textbf{b}-axis of TaIrTe$_4$. \textbf{d}, Photocurrent in TaAs plotted against incident power, at a bias of 100 $\mu$V. Panel \textbf{a} adapted with permission from \cite{ma2019observation}, panel \textbf{b} adapted with permission from \cite{kang2019nonlinear}, panel \textbf{c} adapted with permission from \cite{kumar2021room} and panel \textbf{d} adapted with permission from \cite{chi2017ultra}.} \label{fig12}
\end{figure*}

As a large driving current density is crucial for the observation of nonlinear current, low dimensional systems are advantageous in that they are better able to dissipate heat. Therefore, bilayer WTe$_2$~\cite{ma2019observation} and few-layer WTe$_2$~\cite{kang2019nonlinear} were the first materials to be measured for nonlinear Hall effect. Bilayer WTe$_2$ has a pair of coupled Dirac fermions. The presence of these topological band inversions implies substantial Berry curvature, and the inversion symmetry can be further manipulated by gating field. Operating at 177.77 Hz, as depicted in Fig. \ref{fig12}a, bilayer WTe$_{2}$, which contains only one mirror line in crystal plane, demonstrates a substantial nonlinear Hall current. The amplitude of the transverse voltage response is depicted in Fig. \ref{fig12}b, which shows a peak about 30 $\mu$V and this happens when the current is perpendicular to the mirror plane. The study indicates that both intrinsic factors, such as the Berry curvature dipole, and extrinsic factors like spin-dependent scatterings, collectively contribute to the observed signal in few-layer T$_d$-WTe$_2$. As for the voltage responsivity, bilayer WTe$_2$ reaches 10$^{4}$ V W$^{-1}$ at dc, underscoring its considerable potential for terahertz detection. 

TaIrTe$_{4}$ is a Type-II Weyl semimetal first proposed by density functional theory simulation ~\cite{koepernik2016tairte}, with point group symmetry $C_{2v}$. Owing to its ternary nature with broken inversion symmetry, TaIrTe$_{4}$ hosts substantial band overlapping at the Fermi level compared to its binary counterpart WTe$_2$, and is widely studied for its optical properties. Under exposure to electromagnetic waves, TaIrTe$_{4}$ exhibits a significant dc current and 2$\omega$ current (see Fig. \ref{fig12}c) response \cite{kumar2021room}, particularly at a central frequency of around 2.4 GHz, with the cutoff frequency extending to approximately 5 GHz. In the visible regime, TaAs ~\cite{chi2017ultra} exhibits a good ultra-broadband photoresponse at room temperature as well(Fig. \ref{fig12}d). Its responsiveness spans from blue (438.5 nm) to mid-infrared (10.29 $\mu$m) light wavelengths. With a responsivity of 78 $\mu$A W$^{-1}$, TaAs emerges as a promising candidate.

\begin{figure*}[pos=!h]
	\centering
		\includegraphics[width=1\textwidth]{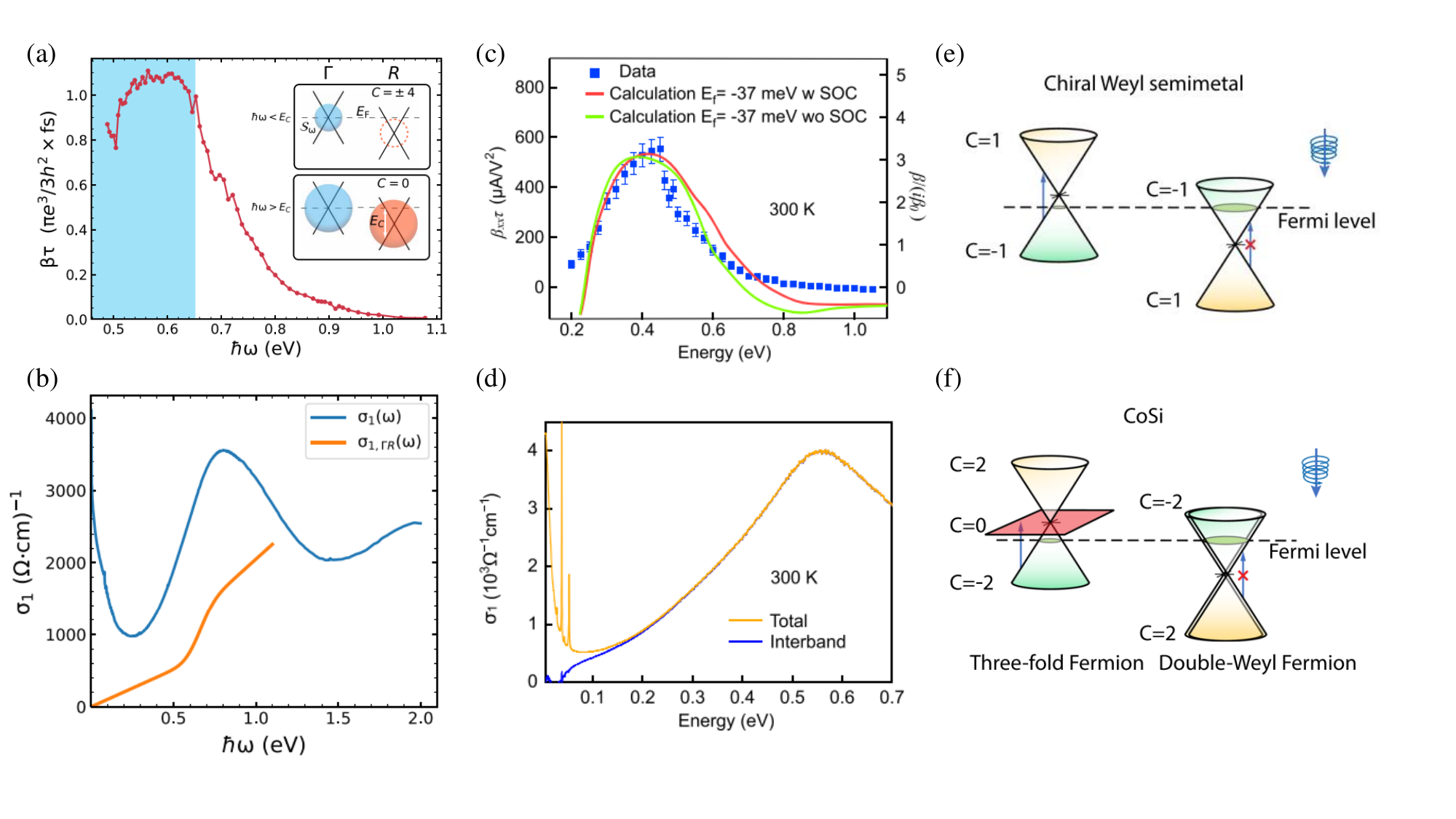}
	  \caption{\textbf{Experiment signatures and theory of topological chiral 
         crystal.} \textbf{a}, CGPE amplitude of RhSi plotted with respect to photon energy showing abrupt quenching when the photon energy exceeds 0.65 eV. The inset schematic in k-space illustrates the surface S${\omega}$, indicating allowed optical transitions for a given photon energy $\hbar\omega < E_C$. Below $E_C$, only a single node is available for transitions and the Berry flux C = $\pm$4; above $E_C$, both nodes with opposite chirality contribute Berry flux and C = 0. \textbf{b}, The blue curve depicts the optical conductivity of RhSi. The peak indicates a scattering time is about 8.6 fs. Meanwhile, the orange line shows the optical conductivity at $\Gamma$ and R nodes. \textbf{c}, Second-order CPGE photo-conductivity of CoSi versus incident photon energy, alongside ab-initio CPGE current calculations with spin-orbit coupling or not at 300 K. \textbf{d}, At 300 K, experimentally obtained total optical conductivity~(gold) and optical conductivity stem from interband transition~(blue) of CoSi. \textbf{e}, At different energies, the chiral Weyl semimetal CoSi will exhibit two Weyl nodes. Both of them are monopoles and they have opposite charges (+1 and -1). \textbf{f}, In the absence of spin-orbit coupling within CoSi, distinct nodes emerge: a threefold node and a double Weyl node, each positioned at varying energy levels and momentum states. The threefold fermions possess Chern numbers of $\pm 2$ and 0. Due to spin degeneracy, the combined charge at the two modes is either positive or negative, totaling $\pm 4$. Under circularly polarized light, excitation around the left Weyl fermion or the threefold fermions generates a CPGE, as the others are Pauli blocked. Panels \textbf{a} and \textbf{b} adapted with permission from \cite{rees2020helicity} and panels \textbf{c} - \textbf{f} adapted with permission from \cite{ni2021giant}.}\label{fig13}
\end{figure*}

Next, we discuss the nonlinear photocurrent in chiral topological materials. While chiral Weyl semimetals were originally proposed to host quantized photocurrent \cite{de2017quantized} as shown in Fig. \ref{fig13}e, the small energy separation between Weyl nodes with opposite chiralities makes it extremely hard to experimentally realize photocurrent quantization. RhSi and CoSi~\cite{rees2020helicity, ni2021giant} stand out due to their multifold fermion dispersion and large energy separation between the pair of multifold nodes, rendering them ideal candidates to exhibit the quantized CPGE \cite{de2017quantized}. In Fig. \ref{fig13}a, the CPGE conductivity $\beta\tau$ of RhSi is measured as a function of incident photon energy, reaching its maximum when the photon energy is approximately 0.65 eV. Beyond interband transitions near the Weyl nodes, additional contributions are observed, as shown in Fig. \ref{fig13}b. 

Although CoSi~\cite{ni2021giant} has a similar band structure as RhSi, its relativity smaller spin orbit coupling and higher sample quality lead to better photocurrent quantization and long relaxation time. The terahertz emission experiments in CoSi demonstrate a substantial longitudinal photo-conductivity at room temperature, peaking at $400$ meV and reaching around 550 $\mu$A/V$^2$~(Fig. \ref{fig13}c,d). For frequencies below $0.6$ eV, all interband transitions in CoSi occur at the nodes $\Gamma$ and $R$, while for frequencies below $0.2$ eV, interband transitions solely occur at the $\Gamma$ point. 

\begin{figure*}[pos=!h]
	\centering
        \includegraphics[width=1\textwidth]{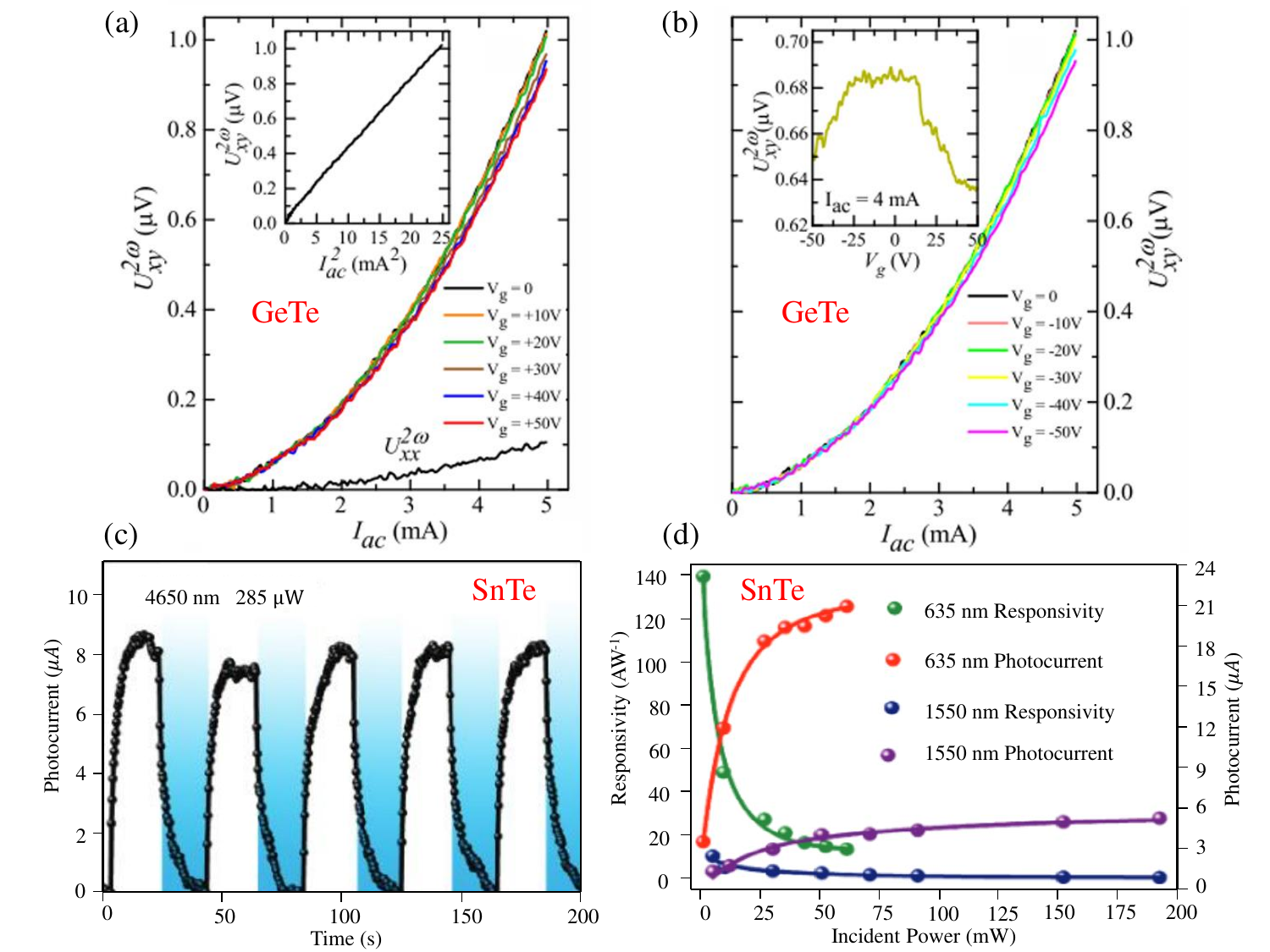}
	  \caption{\textbf{Experiment review for ferroelectric topological materials.} \textbf{a} and \textbf{b}, The transverse second-harmonic voltage ($U_{xy}^{2\omega}$) shows nonlinear Hall effect behavior with varying gate voltages. The inset of Figure (a) shows the linear relation between the transverse voltage and the quadratic current. At zero gate voltage, $U_{xy}^{2\omega}(I^2)$ is strictly linear, while $U_{xx}^{2\omega}$ is much smaller. Positive and negative gate voltages symmetrically affect the curves due to the gate electric field. In the inset of Figure \textbf{b}, $U_{xy}^{2\omega}(V_g)$ is shown for a fixed ac current of $I_{ac} = 4$ mA. \textbf{c}, Time-dependent photocurrent of a SnTe device under 4650 nm wavelength light (excitation power: 285 $\mu$W). The device has a 50 nm channel length, with V$_{SD}$ = 1 V and V$_G$ = 0 V. \textbf{d}, SnTe device's photocurrent and responsivity are plotted respect to incident power at 635 nm and 1550 nm wavelengths. Panels \textbf{a} and \textbf{b} adapted with permission from \cite{orlova2023gate} and panels \textbf{c} and \textbf{d} adapted with permission from \cite{yang2018ultra}.} \label{fig14}
\end{figure*}

On the ferroelectric topological materials side, as previously discussed in the theoretical section, GeTe stands out as a possible ferroelectric topological insulator with point group R3m under alloy engineering. Remarkably, GeTe exhibits a substantial nonlinear Hall effect at room temperature~\cite{orlova2023gate}. Besides that, the response current can be tuned by a gate, leveraging the dependence of the Rashba splitting parameter on external electric fields~\cite{di2012electric, rinaldi2018ferroelectric, orlova2022dynamic}. As shown in Fig. \ref{fig14}a, b, the nonlinear Hall signal reaches approximately 1 $\mu$V at room temperature under varying gate voltages, establishing a correlation between the signal and gate voltages. This outcome introduces both a novel material and a gate-controlled methodology for efficient broadband photodetection. Another ferroelectric topological insulator, SnTe, shows a high responsivity of photocurrent\cite{yang2018ultra}. Fig. \ref{fig14}c illustrates the time-dependent photocurrent due to incident light with a wavelength of 4650 nm, reaching a noteworthy 8 $\mu$A with consistent cycle-to-cycle enhancement. Fig. \ref{fig14}d further explores the relationship between photocurrent and responsivity at incident light wavelengths of 635 nm and 1550 nm. The responsivity peaks at approximately 49.03 AW$^{-1}$ for the 635 nm wavelength and 10.91 AW$^{-1}$ for the 1550 nm wavelength. These findings make SnTe a promising candidate for efficient broadband photodetection.

\begin{figure*}[pos=!h]
	\centering
		\includegraphics[width=1\textwidth]{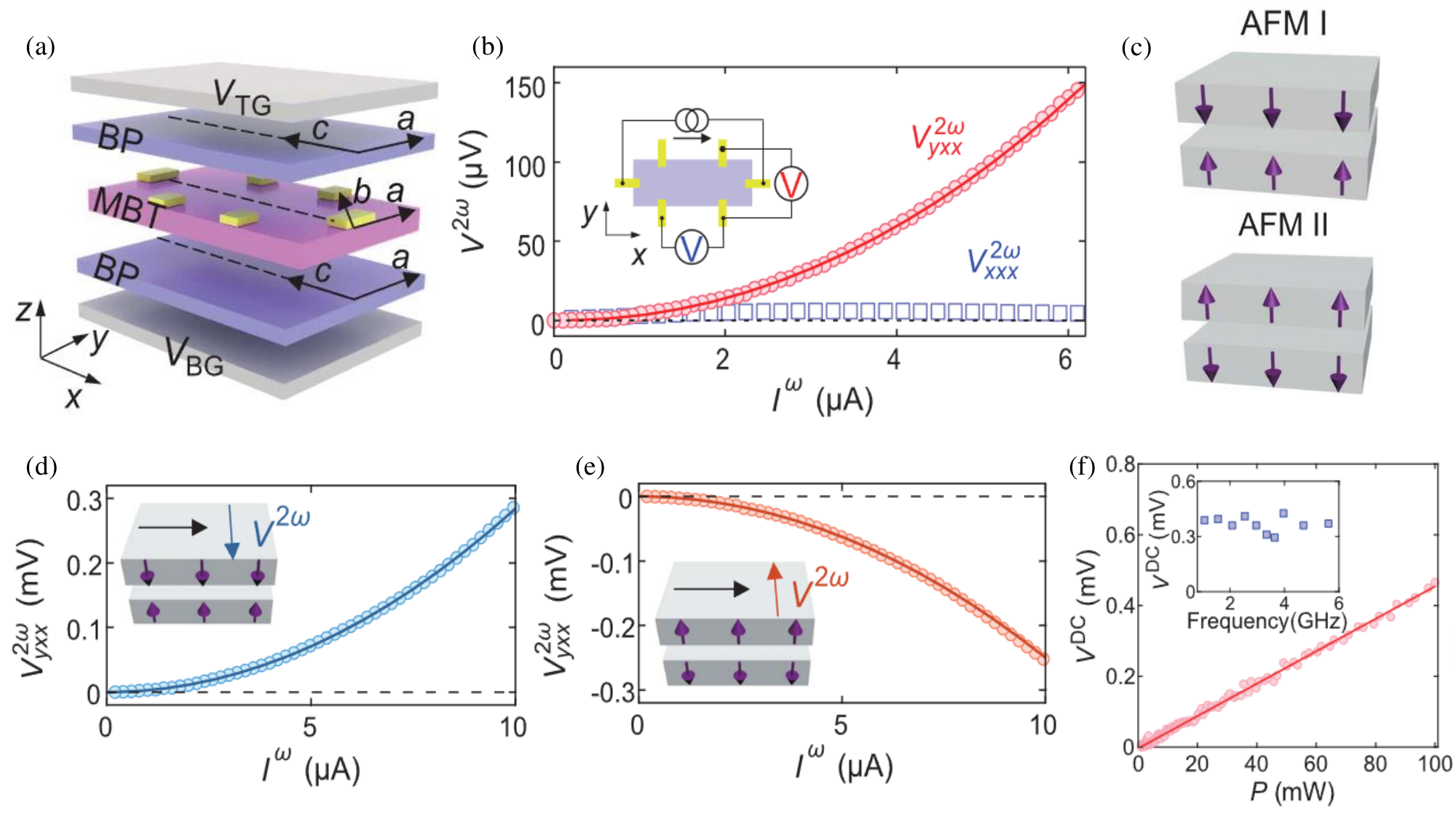}
	  \caption{\textbf{Experimental result for AFM material MnBi$_2$Te$_4$.} \textbf{a}, Schematic illustration of a device comprising a bilayer of black phosphorus (2L BP), followed by a septuple layer (6SL) of MnBi$_2$Te$_4$, and another bilayer of black phosphorus (2L BP). In this context, a septuple layer (SL) of MnBi$_2$Te$_4$ is arranged as Te-Bi-Te-Mn-Te-Bi-Te, with aligned crystalline $a$ axes for both the black phosphorus layers and the MnBi$_2$Te$_4$ layer. \textbf{b}, The longitudinal ($V_{xxx}^{2\omega}$) and Hall ($V_{yxx}^{2\omega}$) voltage. \textbf{c}, Two opposite antiferromagnetic (AFM) states are generated by sweeping $B_z$ from -8 T to 0 T (AFM I) or from +8 T to 0 T (AFM II) in the presence of $E_z$ = -0.17 V/nm. This method is from~\cite{gao2021layer}. \textbf{d}, Nonlinear Hall voltage plotted against incident current for AFM I. \textbf{e}, Nonlinear Hall voltage plotted against incident current for AFM II. \textbf{f}, Microwave rectification measured through intrinsic nonlinear Hall effect (NHE). Inset: Direct current (DC) signal plotted against microwave frequency. This figure adapted with permission from \cite{gao2023quantum}.} \label{fig15}
\end{figure*}

Beyond time reversal symmetric nonmagnetic materials, magnetic materials also exhibit strong nonlinear response signals in several experiments. A notable example is MnBi$_2$Te$_4$, a two-dimensional antiferromagnetic (AFM) material that supports axion electrodynamics. Recent work on the heterostructure of the two-dimensional MnBi$_2$Te$_4$ layers has revealed a large nonlinear current related to time reversal symmetry breaking~\cite{gao2023quantum}. Unlike the previously mentioned materials, the origin of this nonlinear current lies in the quantum metric dipole~\cite{gao2014field, wang2021intrinsic, liu2021intrinsic, lahiri2022intrinsic, smith2022momentum}. The quantum metric is defined as the real part of the quantum geometric tensor: $g_{\alpha\beta}=Re\sum_{m\neq n}[\braket{\mu_n|i\partial_{k_\alpha}\mu_m}\braket{\mu_m|i\partial_{k_\beta}\mu_n}]$, which quantifies the distance between neighboring Bloch wave functions. The device structure of MnBi$_2$Te$_4$ is depicted in Fig. \ref{fig15}a, and here black phosphorus (BP) is to break $C_{3z}$ rotational symmetry. Here their $a$ axes are aligned to guarantee the nullification of Berry curvature and Berry curvature dipole, and as a result the $\mathcal{PT}$ symmetry is not broken. The resulting response signal is presented in Fig. \ref{fig15}b, where under 1.8 K, the transverse nonlinear response $V_{yxx}^{2\omega}$ reaches 150 $\mu V$ with Hall dominance. Further investigation reveals a correlation between the nonlinear signal and the opposite AFM states, which are shown in Fig. \ref{fig15}c. AFM I exhibits a prominent positive nonlinear voltage~(Fig. \ref{fig15}d), while AFM II shows a flipped sign compared to AFM I~(Fig. \ref{fig15}e). This result provides additional evidence that the nonlinear Hall current is from the quantum metric dipole, since it is expected to be odd under time reversal operations between two AFM states. Additionally, MnBi$_2$Te$_4$ exhibits a broadband response in rectified DC voltage, encompassing WiFi frequencies within the range 2.4 and 5 GHz~(Fig. \ref{fig15}f). Consequently, magnetic topological materials emerge as strong candidates for broadband photodetection, with their origin rooted in the quantum metric dipole rather than Berry curvature and Berry curvature dipole. However, the regulation of electronic motions by the quantum metric remains largely unknown, and examples of such materials are rare. Therefore, further exploration of material candidates and underlying mechanisms is warranted.

In addition to the aforementioned materials, we summarize the experimentally studied materials for nonlinear response in Table 2, together with various experimental conditions.

\begin{table*}[width=1\textwidth, pos=!h]
\caption{Experiments on the nonlinear Hall effect. This format of this table refers to \cite{du2021nonlinear}.}\label{Tab:Exp}
\begin{tabular*}{\tblwidth}{@{}LLLLLLL@{}}
\toprule
  Materials & Dimension & \makecell[c]{Temperature\\~(K)} & \makecell[c]{Input current\\frequency\\~(Hz)} & \makecell[c]{Input current\\maximum\\~($\mu$A)} & \makecell[c]{Output voltage\\maximum\\~($\mu$V)} & \makecell[c]{Carrier density\\~(cm$^{-2}$) in 2D\\
    ~(cm$^{-3}$) in 3D} \\
\midrule
 Bilayer WTe$_{2}$ ~\cite{ma2019observation} & 2 & 10-100 & 10-1000 & 1 & 200 & $\sim10^{12}$\\
      BP/MnBi$_{2}$Te$_{4}$ ~\cite{gao2023quantum} & 2 & 1.8-30 & $\omega$ & 10 & 350 & $\sim10^{12}$\\
      Few-layer WTe$_{2}$ ~\cite{kang2019nonlinear} & 2 & 1.8-100 & 17-137 & 600 & 30 & $\sim10^{13}$\\
      Twisted bilayer WSe$_{2}$ ~\cite{huang2023giant} & 2 & 1.5-30 & 4.579 & 0.04 & 20,000 & $\sim10^{12}$\\
      Twisted bilayer graphene ~\cite{duan2022giant} & 2 & 1.7-80 & 17.777 & 1 & 600 &-\\
      4SL-MnBi$_{2}$Te$_{4}$ ~\cite{wang2023quantum} & 2 & 1.8-30 & 17.777 & 10 & 200 & $\sim10^{12}$\\
      1T-CoTe$_{2}$ ~\cite{hu2023terahertz} & 2 & 77-300 & 0.02T-0.12T & - & 2 & -\\
      Strained monolayer WSe$_{2}$ ~\cite{qin2021strain} & 2 & 50-140 & 17.777 & 4.5 & 20 & $\sim10^{13}$\\
      Bi$_{2}$Se$_{3}$ surface ~\cite{he2021quantum} & 2 & 2-200 & 9-263 & 1,500 & 20 & $\sim10^{13}$\\
      Corrugated bilayer graphene ~\cite{ho2021hall} & 2 & 1.5-15 & 77 & 0.1 & 2 & $\sim10^{12}$\\
      hBN/graphene/hBN ~\cite{he2022graphene} & 2 & 1.65-200 & 31 & 5 & 100 & $\sim10^{11}$\\
      GeTe ~\cite{orlova2023gate} & 2 & 300 & $\omega$ & 5,000 & 1 & -\\
      LaAlO$_{3}$/SrTiO$_{3}$ ~\cite{lesne2023designing} & 2 & 1.5-30 & $\omega$ & 200 & 120 & -\\
      CoSi ~\cite{esin2021nonlinear} & 2 & 1.2-4.2 & 7,700 & 500 & 1 & -\\
      Pb$_{1-x}$Sn$_{x}$Te ~\cite{nishijima2023ferroic} & 2 & 5-300 & 1000 & 100 & 1000 & -\\
      Twisted trilayer graphene ~\cite{zhang2022diodic} & 2 & 0.02 & $\omega$ & 0.03 & 0.3 & -\\
      Fe$_{3}$GeTe$_{2}$ ~\cite{esin2022second} & 2 & 1.4-4.2 & $\omega$ & 4,500 & 0.25 & -\\
      Ce$_{3}$Bi$_{4}$Pd$_{3}$ ~\cite{dzsaber2021giant} & 3 & 0.4-4 & 0-13 & 10,000 & 0.8 & $\sim10^{19}$\\
      T$_{d}$-TaIrTe$_{4}$ ~\cite{kumar2021room} & 3 & 2-300 & 13.7-213.7 & 600 & 120 & $\sim10^{19-20}$\\
      BaMnSb$_{2}$ ~\cite{min2023strong} & 3 & 200-400 & 17.777-117.777 & 200 & 400 & $\sim10^{18-21}$\\
      $\alpha$-~(BEDT-TTF)$_{2}$I$_{3}$ ~\cite{kiswandhi2021observation} & 3 & 4.2 & d.c. & 1,000 & 8-9 & $\sim10^{17}$\\
      T$_{d}$-MoTe$_{2}$ ~\cite{tiwari2021giant} & 3 & 2-40 & 17-277 & 5,000 & 40 & $\sim10^{19-20}$\\
      Bulk WTe$_{2}$ ~\cite{shvetsov2019nonlinear} & 3 & 1.4-4.2 & 110 & 4,000 & 2 &-\\
      Cd$_{3}$As$_{2}$ ~\cite{shvetsov2019nonlinear} & 3 & 1.4-4.2 & 110 & 4,000 & 1 &$\sim10^{18}$\\
      PbTaSe$_{2}$ ~\cite{itahashi2022giant} & 3 & 2-300 & 13 & 4,000 & 3,200 & $\sim10^{22}$\\
     ~(Pb$_{1-x}$Sn$_{x}$)$_{1-y}$In$_{y}$Te ~\cite{zhang2022giant} & 3 & 3-20 & 13.333 & 100 & 40 & $\sim10^{15-18}$\\
\bottomrule
\end{tabular*}
\end{table*}

\subsubsection{Difference between 2D and bulk materials}

It is essential to highlight the distinctions between 2D materials and bulk materials. There exist three principal differences between these two categories:~(1) While the nonlinear Hall effect in both 2D and bulk materials can be traced back to the Berry curvature dipole, 2D materials, especially high mobility graphene and semimetals, exhibit an additional contribution to nonlinear Hall current: skew-scattering of electrons by impurities or phonons. Skew-scattering involves the deflection of electrons by scattering centers, influenced by their spin or valley degrees of freedom. This process results in the generation of a transverse current, proportional to the square of the longitudinal current~\cite{du2021nonlinear}.~(2) In bulk materials, the nonlinear Hall effect is typically less pronounced compared to 2D materials. This discrepancy arises from the smaller Berry curvature dipole and the limitation of longitudinal current density imposed by small resistance in bulk topological materials. In contrast, 2D materials benefit from the enhancement of the Berry curvature dipole through quantum confinement, and the longitudinal current can be augmented by elevating electron mobility~\cite{li2021nonlinear}.~(3) The nonlinear Hall effect in bulk materials is inherently determined by intrinsic properties such as band structure, crystal symmetry, and Fermi level. On the contrary, 2D materials offer greater flexibility in allowing the application of external tuning knobs. By manipulating 2D materials through double gates~\cite{zhang2018electrically}, adjustments to band structure, Fermi level, and carrier density become feasible, providing more avenues to enhance responsivity.

\begin{figure*}[pos=!h]
	\centering
		\includegraphics[width=1\textwidth]{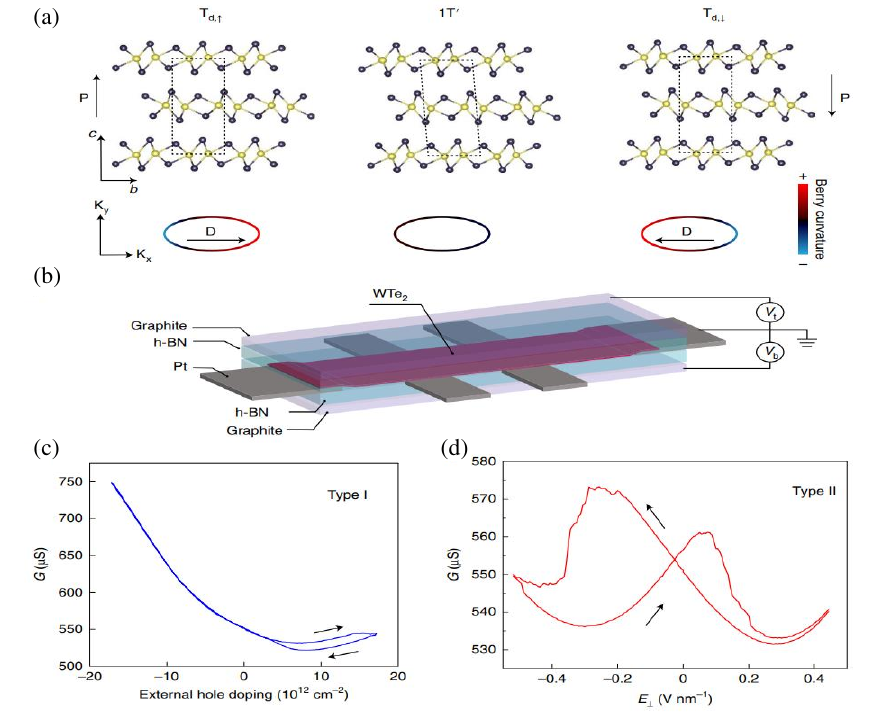}
	  \caption{\textbf{Berry curvature manipulation in few-layer WTe$_2$.} \textbf{a}, The T$_{d,\uparrow}$, 1T$'$, and T$_{d,\downarrow}$ stacking configurations of WTe$_2$ and the distribution of Berry curvature, showing that the T$_{d,\uparrow}$ and T$_{d,\downarrow}$ configurations generate non-trivial Berry curvature dipoles. \textbf{b}, The WTe$_2$ device is sandwiched between two layers of hexagonal boron nitride (hBN) of thickness $d_{b}$ and $d_{t}$, followed by two layers of graphene, which serve as the electrodes. \textbf{c} and \textbf{d} In a five-layer sample, the electrical conductance $G$ undergoes two different hysteresis shapes under different doping regimes. Under condition $V_{t}/d_{t}=V_{b}/d_{b}$, the hole doping regime, the conductance goes through a rectangular shaped hysteresis. \textbf{d}, Under condition $-V_{t}/d_{t}=V_{b}/d_{b}$, $G$ exhibits a butterfly-shape hysteresis~(type II). Positive $E_{\perp}$ is defined along the +c axis. This suggests we can engineer a phase transition by changing the gate voltage. This figure adapted with permission from \cite{xiao2020berry}.} \label{fig16}
\end{figure*}

Moreover, 2D materials exhibit significant potential in the dynamic control of band structure and Berry curvature, presenting a distinct advantage due to the strong correlation between photocurrent and the quantum wavefunctions. The ability to manipulate Berry curvature via stacking order, twist angle, and relative inter-layer shifting, opens avenues for regulating nonlinear response, a crucial aspect in advancing electronic applications such as photodevices and memory cells. A key piece of experimental evidence demonstrating manipulation of Berry curvature has been demonstrated through electrically driven stacking transitions in WTe$_2$~\cite{xiao2020berry}. Theoretical investigations have delved into the unique stacking orders of WTe$_2$, revealing three hidden phases with distinct symmetries~\cite{kim2017origins, yang2018origin}: the 1T' structure (non-polar monoclinic), T$_d$ phase (polar orthorhombic) with T$_{d,\uparrow}$ (upward spontaneous polarization), and T$_{d,\downarrow}$ (downward spontaneous polarization). Figure. \ref{fig16}a illustrates that these phases share identical single-layer configurations, differing only in the relative interlayer position along the b-axis. The relative position of the layers determine the Berry curvature distribution, leading to a Berry curvature dipole (Fig. \ref{fig16}a). Therefore, WTe$_2$ is ideal for demonstrating dynamically tunable stacking transitions, which can be electrically read out by Berry curvature related nonlinear Hall current. For experiments, the fabricated device and results are shown as Fig. \ref{fig16}b,c,d. 

\begin{figure*}[pos=!h]
	\centering
		\includegraphics[width=1\textwidth]{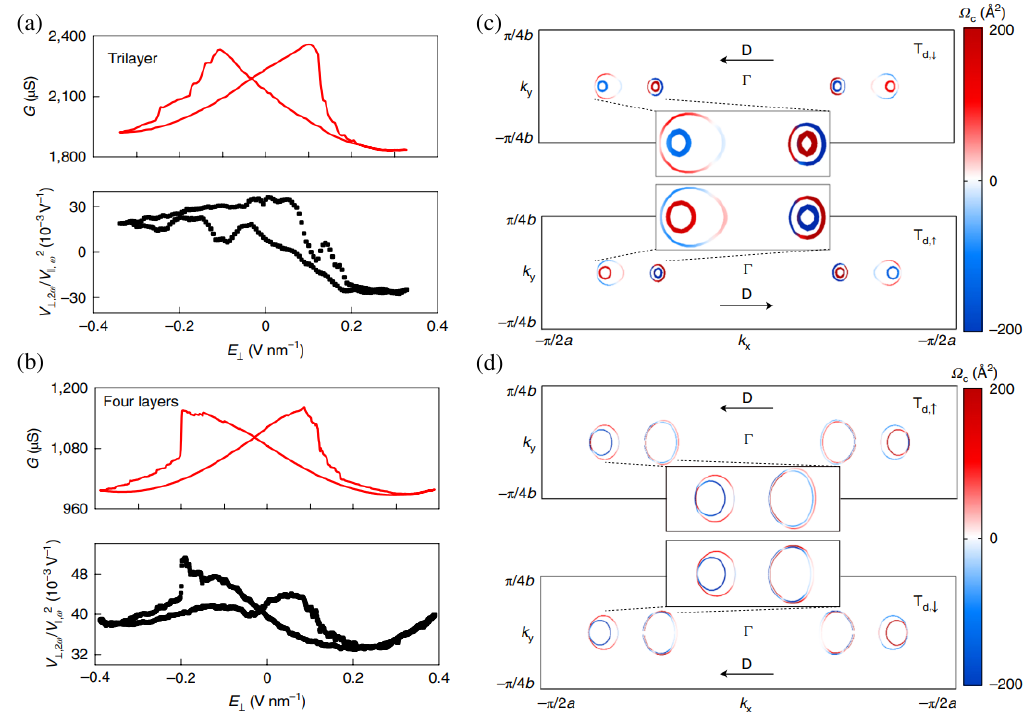}
	  \caption{\textbf{Berry curvature behaviour in $T_{d,\uparrow}$ to $T_{d,\downarrow}$ process.} \textbf{a} and \textbf{b}, $G$ (up) and nonlinear Hall signal (low) for trilayer and tetralayer WTe$_2$. While both exhibit butterfly-shaped hysteresis in longitudinal conductance, only the trilayer shows a reversed sign in the nonlinear Hall signal, indicating a sign flip of the Berry curvature dipole. \textbf{c} and \textbf{d}, The distribution of Berry curvature ($\Omega_c$) for $T_{d,\uparrow}$ and $T_{d,\downarrow}$ phases in trilayer and tetralayer WTe$_2$. It shows how Berry curvature reverses for trilayer while remaining invariant for tetralayer, indicating the presence of nonlinear Hall current and Berry curvature memory. This figure adapted with permission from \cite{xiao2020berry}.} \label{fig17}
\end{figure*}

The memory behaviour of Berry curvature and Berry curvature dipole can be realized by phase transition driven by the electric field. The mechanics of these transitions are demonstrated by measuring nonlinear signal. In the transition between $T_{d,\uparrow}$ and $T_{d,\downarrow}$, both trilayer and tetralayer WTe$_2$ exhibit clear hysteresis in G and the nonlinear Hall signal $V_{\perp,2\omega}/(V_{\parallel,\omega})^2$, as shown in Fig. \ref{fig17}a, b. Notably, The nonlinear Hall signal reverses its sign in trilayer WTe$_2$, indicating a reversal in the direction of the Berry curvature dipole, because the nonlinear Hall signal is proportional to the Berry curvature dipole strength. In contrast, the nonlinear Hall signal remains unchanged in tetralayer WTe$_2$. The calculations of Berry curvature for trilayer and tetralayer WTe$_2$, as depicted in Fig. \ref{fig17}c, d, validate that the sign change is related with the number (parity) of layers, revealing that Berry curvature inverts its direction in odd layers while maintaining the same sign in even layers during stacking order transitions. These findings underscore the dynamic control of Berry curvature through stacking-order transitions, providing a means to adjust nonlinear response for improved photodetection and memory devices.

\section{Application in broadband electromagnetic wave detection}

\begin{figure*}[pos=!h]
	\centering
		\includegraphics[width=1\textwidth]{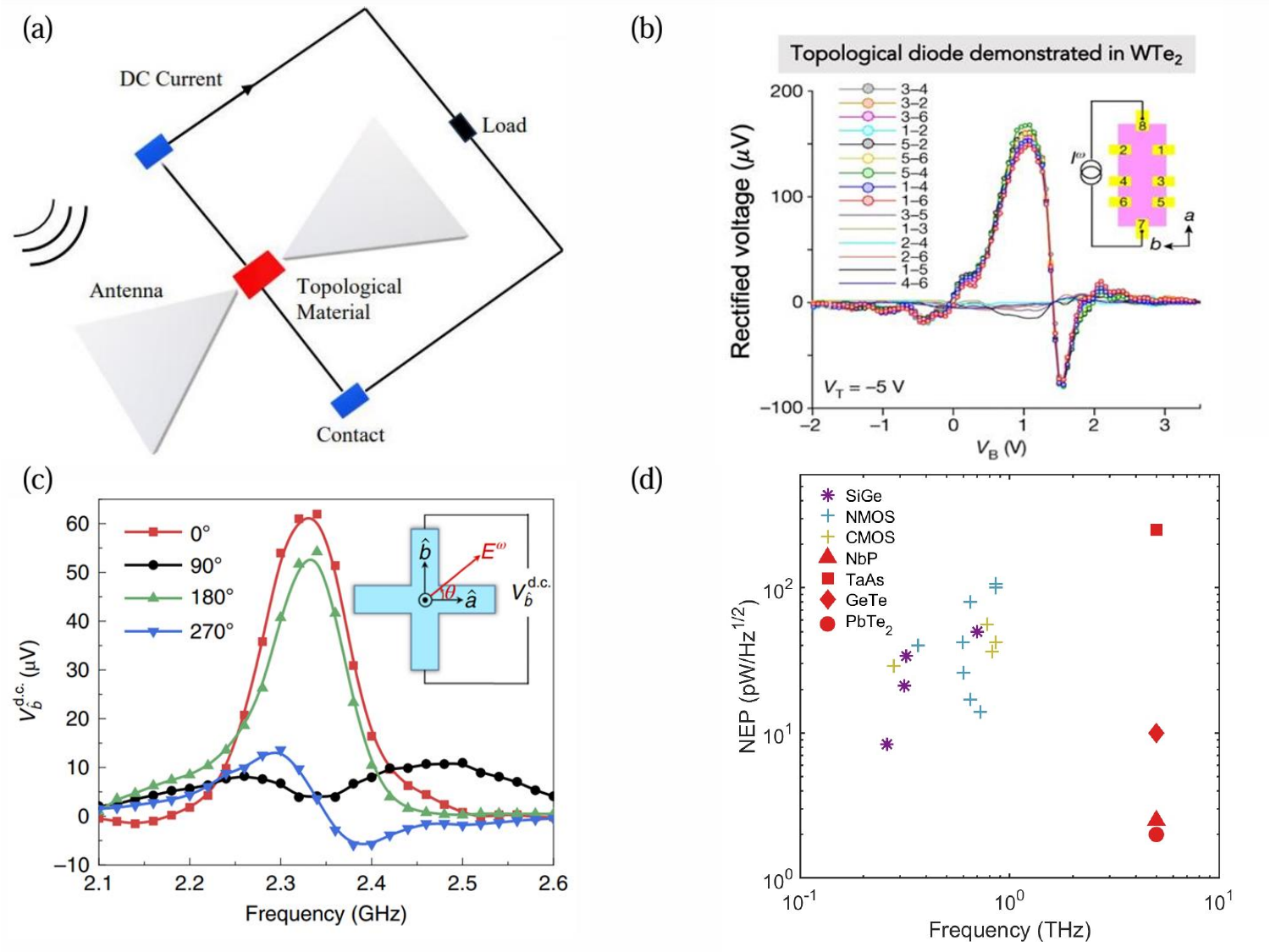}
	  \caption{\textbf{Application to photodetection I.} \textbf{a}, Possible device for self-powered room temperature infrared sensing and Ultra-efficient harvesting of ambient RF energy. \textbf{b}, Rectified voltage in WTe$_{2}$, showcasing its capability to efficiently harvest extremely low power radio frequency (RF) signals with a high responsivity. \textbf{c}, Demonstration of rectification utilizing the nonlinear Hall effect in TaIrTe$_4$, exhibiting excellent detectivity particularly at approximately 2.4 GHz. \textbf{d}, The NEP and frequency of topological photodetection, clearly showing that the device can surpass the existing devices in energy-efficiency~(zero-bias and room temperature). This figure summarized specially for silicon-based high-frequency electronic device. The data for \textbf{d} is taken from \cite{hillger2018terahertz}. Panel \textbf{b} adapted with permission from \cite{ma2019observation} and panel \textbf{c} adapted with permission from \cite{kumar2021room}.}
    \label{fig18}
\end{figure*}

Conventional diodes rely on built-in electric fields within confined p-n junctions to rectify AC current into DC current, finding applications in diverse fields such as power transmission, high frequency rectification and photodetection. The fundamental working principles of topological diodes involve interband photocurrent and intraband transitions induced nonlinear Hall current. These processes rectify the ac electric field of circular or linear polarized light into a direct current. The conceptual structure of an application device is depicted in Fig. \ref{fig18}a. Compared to conventional diodes, the topological diode harnesses intrinsic quantum dipoles carried by itinerant electrons, eliminating the need for junctions or built-in electric fields. This innovation could be used in a broad range of topological semiconductors, metals, and superconductors, offering significant potential for building ultra-sensitive and energy-efficient devices. 

Several key highlights for high frequency rectification based on topological diodes include:~(1) Topological diode devices exhibit exceptional efficiency in detecting and harvesting electromagnetic waves across the infrared, THz, and WiFi frequencies, especially in the ultra-low power density regime. This efficacy arises from their robust coupling with topological electrons.~(2) The active area of these devices encompasses the entire material, as the effect is an intrinsic property of homogeneous materials rather than relying on heterojunctions. This characteristic enhances the versatility and simplicity of device design.~(3) Topological diodes have the capacity to surpass certain limits imposed by traditional p-n junction mechanics, particularly in realms such as low-power rectification and sensing.~(4) These diodes demonstrate resilience by maintaining functionality under ambient conditions, and across both macro-scale and nano-scale materials. Their robust topological nature also makes them highly resistant to impurities and external perturbations.~(5) Topological diodes work at zero bias, requiring no external voltage for their functioning, and they can be self-powered. These highlights collectively advocate for the development of revolutionary energy-efficient devices. In the subsequent discussion, we will delve into two primary potential applications for broadband rectification in the terahertz and radio frequency range, both of which have been experimentally demonstrated.

\begin{figure*}[pos=!h]
	\centering
		\includegraphics[width=1\textwidth]{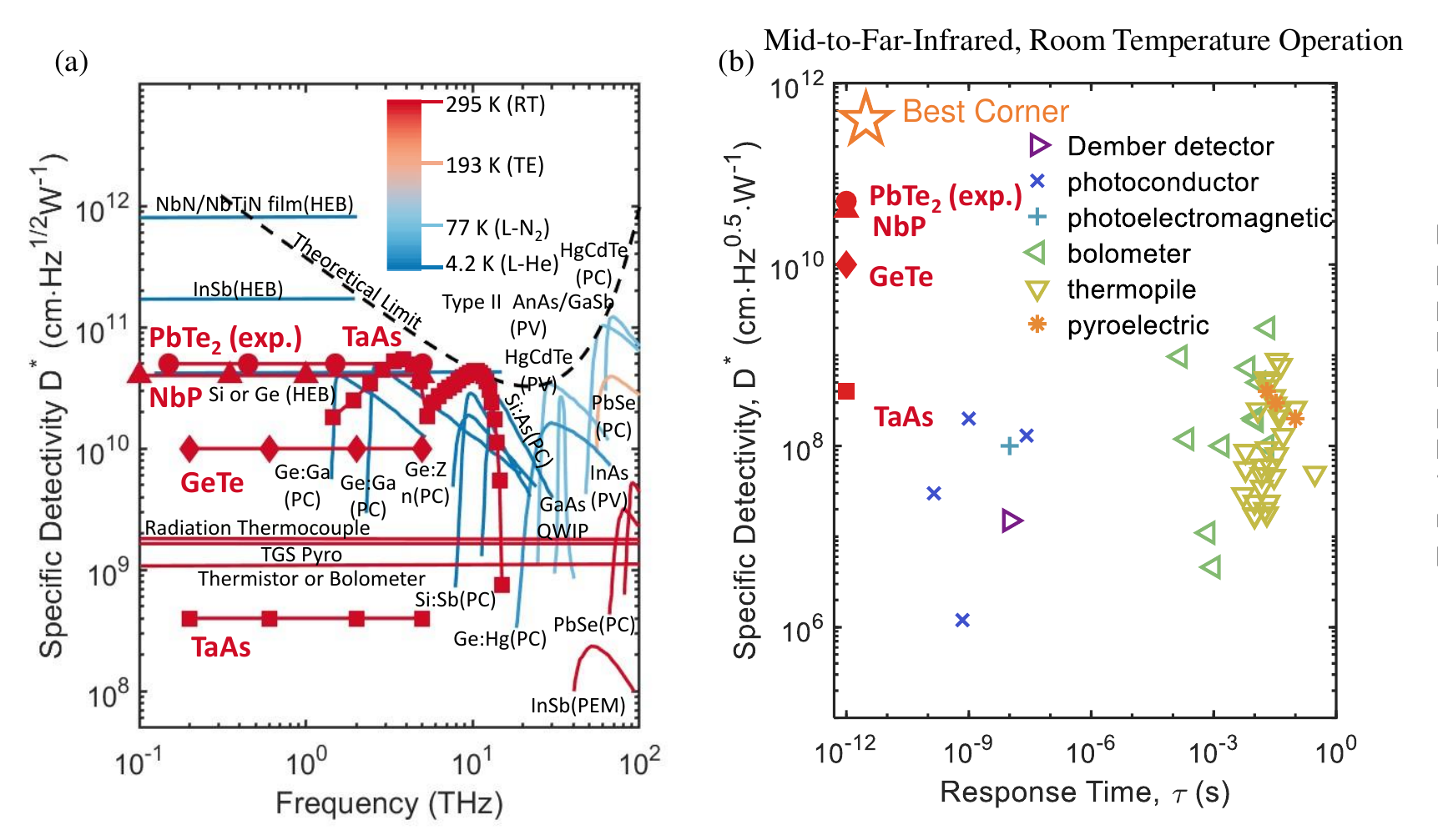}
	  \caption{\textbf{Application to photodetection II.} \textbf{a} and \textbf{b}, Detectivity of topological photodetection with respect to frequency and response time at room temperature, clearly showing the device can surpass the existing devices in detectivity and response speed. \textbf{a}, The data are colored according to the operation temperature. Abbreviations: RT - room temperature; TE - thermoelectric cooling; L-N$_2$ - liquid nitrogen cooling; L-He - liquid helium cooling; exp. - experiments; PC - photoconductor; PV - photovoltaic; PEM - photoelectromagnetic; HEB - hot electron bolometer; TSG Pyro - triglycine sulfate pyroelectric detector. \textbf{b}, All the data are detectors working at room temperature and the spectral ranges are mid-to-far infrared. The data for \textbf{a} is taken from \cite{rogalski2012history, rogalski2012progress, gunapala1994high, kozlowski1997performance, rogalski2003infrared} and the data for \textbf{b} is taken from \cite{graf2007review, chi2005characterization, piotrowski2004uncooled, mohseni1999uncooled, lee1998exploration, kim1998investigation, wang2004ir, yildiz2004microbolometers, lee1999high, tezcan2003low, yoneoka2011ald, rogalski2002infrared, almasri2001self, sedky1998characterization, lattanzio2011complementary, ignatiev1998pyroelectric, jerominek1996micromachined, muralt2001micromachined}.} \label{fig19}
\end{figure*}

\subsection{Application for terahertz sensing}

Terahertz radiation's safety advantage—non-ionizing and non-invasive nature—underpins its critical role in applications like night vision, tumor diagnostics in medical imaging, 6G wireless tech (using frequencies up to 0.3 THz), and advanced human-computer interfaces based on terahertz imaging. However, detecting waves within the 0.3 THz to 30 THz frequency range has historically been challenging due to the low photon energy, making direct bandgap absorption in most semiconductor devices impractical. Approaching from the microwave side, conventional high-frequency electronics also struggle with impedance mismatches, signal reflections, and diffusion time through P-N barriers, which greatly increases manufacturing costs for circuits operating beyond 0.3 THz. Overcoming these challenges necessitates the development of highly efficient rectification devices for THz sensing. 

%Topological diode devices emerge as strong candidates owing to several merits:~(1) 
Unlike most semiconductors, topological materials, with zero or very narrow bandgaps ($\sim10$ meV) and linear dispersion, strongly interact with terahertz waves in the desired frequency range. The response dc current in topological diode devices is directly proportional to Berry curvature or Berry connection, which can serve as direct guiding principles for computational material design. On the practical application side, thermal fluctuations in the nonlinear current are the dominant noise source for determining noise equivalent power (NEP) around room temperature. The large current responsivity in topological materials leads to a promising NEP of approximately 2.5 pW/Hz$^{1/2}$ even at 300 K~\cite{zhang2021terahertz}. And the response time is limited only by the carrier scattering time, at the level of picoseconds, making it significantly faster than existing room-temperature sensing techniques. Fig. \ref{fig18}d and Fig. \ref{fig19} illustrate realistic calculations of detectivity, response time and NEP for different frequency and different topological materials, affirming their potential to overcome existing challenges and serve as the foundation for innovative photodetectors. Specifically, the Dirac semimetal Ir$_{0.7}$Pt$_{0.3}$Te$_2$ with a metal-Ir$_{0.7}$Pt$_{0.3}$Te$_2$-metal-based room-temperature photodetector demonstrates superior performance. At 0.12 THz, it achieves a responsivity of 0.52 A/W with a response time of 1-3 $\mu$s. At 0.3 THz, the responsivity is 0.45 A/W, and the NEP is less than 24 pW/Hz$^{1/2}$~\cite{xu2021colossal}. Moreover, topological quantum materials are sensitive to light polarization, introducing new sensing functionalities beyond photodiode based sensing.~\cite{ma2022intelligent}.

\subsection{Application for radio wave energy harvesting}
Another spectral range of large industry interest is radio frequency (from 3 kHz to 300 GHz), especially the GHz range for WiFi technology. 
WiFi signals have become omnipresent, providing local network and internet access. It would be highly advantageous if electronic devices could directly tap into the radiation within the WiFi bands (2.4 GHz and 5.0 GHz) for wireless charging and power transmission. Despite advancements in technology, conventional diode-based harvesters struggle to meet demands in the GHz frequency range. These harvesters prove inefficient when the power density of electromagnetic (EM) radiation is low, a common scenario with ambient EM radiations (10 nW/cm$^2$). However, the emerging technology leveraging the topological diode effect and devices based on novel topological quantum materials hold the potential to revolutionize RF harvesters for low-energy applications.

In terms of weak signal rectification, numerous advantages underscore the potential of topological diodes for RF energy harvesting:~(1) P-N junction rectifiers are general constrained by thermal voltage at low input power levels, while topological diodes exhibit intrinsic quadratic I-V curves~(Fig. \ref{fig1}b,d). This characteristic eliminates the need for a minimum input power, crucial for harnessing ubiquitous energy from ambient WiFi signals.~(2) Topological rectifiers have demonstrated substantial responsivity~\cite{ma2019observation}, achieving response levels of a few A/W at room temperature even in extremely low-power conditions~($<$10 nW/cm$^2$).~(3) Topological rectifiers also boast lower resistance compared to traditional photodiodes, facilitating easier impedance matching with external antennas. Additionally, integrating a resonant circuit connected to the quantum rectifier can further increase the output voltage.~(4) Base topological materials can be grown at low temperatures directly on fully fabricated silicon chips, streamlining the heterogeneous integration of these devices. WTe$_2$ in Fig. \ref{fig18}b demonstrates significant response voltage in the low-energy range, showcasing its potential for RF harvesting. Furthermore, TaIrTe$_4$ (Fig. \ref{fig18}c) exhibits substantial responsivity around 2.4 GHz, with a cutoff frequency of approximately 5 GHz, covering the WiFi channels~\cite{kumar2021room}.

\section{Outlook}

Nonlinear photocurrent in topological quantum materials is an emerging research area that is expected to have profound impacts on both the fundamental research as a multiphysics diagnostic of quantum materials, and the practical applications in broadband photodetection. A series of experiments have demonstrated superior performance over conventional p-n junctions and thermal bolometers in several key figures of merit. This includes superior responsivity, faster response times, and an expanded detection spectral range. In addition to the photodetection for radiation strength, topological materials can be sensitive to light polarization and helicity, as shown in the TaAs-based polarimeter\cite{tian2022weyl}.

Harnessing nonlinear photocurrent for broadband photodetection holds great promise, unlocking avenues for optimizing high-performance optoelectronic devices. Yet, challenges persist. On the experimental side, nonlinear topological effects compete with other phenomena in the material. Defects such as impurities or domains can break the symmetries required for topological phenomena. Therefore, the development of synthetic approaches of topological materials in large area and with good material quality is very critical for scaling up this technology. Metal flux and chemical vapor transport (CVT) approaches have been the two major methods to produce high-quality bulk topological crystals \cite{kumar2020topological, xu2019crystal}, while sputtering approach has been more recently reported for thin films \cite{zhang2023robust, wang2019structural, ding2021switching, zhuang2022large}. Further studies should focus on wafer-scale synthesis of topological materials with controlled thickness and reduced defects. Second, the contacts may also break symmetries or introduce work functions that inhibit detection of topological effects. Because of this, fine control of material synthesis, device fabrication and measurement techniques are needed to observe nonlinear phenomena in topological materials \cite{ma2021nonlinearopticstopologygeometry}. Third, comprehensive material and device characterization approaches for the band structures and transport properties of semimetals are required to fully exploit the physical phenomena and optimize the device performance. Angle-resolved photoemission spectroscopy (ARPES), scanning tunneling microscopy (STM), magneto-optical transport, and quantum oscillation measurements are the commonly used tools \cite{borisenko2014experimental, armitage2018weyl, liu2014discovery, xing2020surface, hu2018quantum}. These measurements should be further developed for high-throughput applications, which is essential for large-scale manufacturing. Finally, thermal budget, chemical compatibility, and mechanical strength of the heterogeneous interfaces should be considered carefully when integrating topological materials with conventional silicon-based technologies.

On the theoretical side, the intricate concept of topology poses a hurdle for semiconductor manufacturing, microelectronics, and optoelectronics communities, confining valuable findings to the fundamental science sector. This gap between theory and application is amplified by complex physical mechanisms and a lack of comprehensive performance analysis. The adaptability of standard device architectures and integration principles is questioned due to this novel mechanism, hindering further investigations into environmental and economic impacts.  Moreover, the predominant focus on fundamental science hampers large-scale growth of topological materials for practical use. Bridging this fundamental-to-application gap is crucial for wider adoption. Therefore, continued research on nonlinear photocurrent in quantum materials and its applications is essential. Resolving these challenges will not only enhance our understanding of quantum materials but also unlock transformative technological potential. %Collaborative efforts are vital for turning theoretical promise into practical applications.  

\section*{Acknowledgments}
We are grateful to Qiong Ma and Liang Fu for helpful discussions. Y. S. and L. P. were supported by the National Science Foundation Materials Research Science and Engineering Center program through the UT Knoxville Center for Advanced Materials and Manufacturing~(DMR-2309083). Y. Z. is supported by the start-up fund at University of Tennessee Knoxville. T.D.N and Y.C.L. acknowledge the partial financial support from Texas A\&M University and US Army Research Office (ARO), under Award No. W911NF2120059.

% Uncomment and use as the case may be
%\begin{theorem} 
%\end{theorem}

% Uncomment and use as the case may be
%\begin{lemma} 
%\end{lemma}


%\bibliographystyle{cas-model2-names}

\bibliographystyle{unsrt}
%\bibliography{main.bbl}
\end{CJK}
\end{document}